\documentclass[12pt]{article}
\pdfoutput=1 

\usepackage[utf8]{inputenc}
\usepackage{cancel,slashed}
\usepackage{bbm}
\usepackage{amssymb}
\usepackage{amsfonts}
\usepackage{amsmath}
\usepackage{graphicx}
\usepackage{cite}

\usepackage{latexsym}
\usepackage{color}
\usepackage{xypic}
\usepackage{cancel,slashed}
\usepackage[linktocpage]{hyperref}
\usepackage{color}
\usepackage{transparent}
\usepackage{footmisc}

\makeatletter 
\def\@xfootnote[#1]{%
  \protected@xdef\@thefnmark{#1}%
  \@footnotemark\@footnotetext}
\makeatother

\def\be{\begin{equation}}
\def\ee{\end{equation}}
\def\bea{\begin{eqnarray}}
\def\eea{\end{eqnarray}}

\newcommand{\PE}{\mbox{PE}}
\newcommand{\rk}{\mbox{rk}}
\usepackage{multirow}
\def\tr{{\rm tr \,}}

\def\sm2{{\mbox{\small 2}}}

\usepackage{tikz}
\usetikzlibrary{arrows,snakes,backgrounds}
\usetikzlibrary{shapes.misc}
\usetikzlibrary{decorations.pathreplacing}

\tikzstyle{gauge}=[circle,draw=blue!50,fill=blue!20,thick, inner sep=0pt,minimum size=1cm]
\tikzstyle{circ}=[circle,draw,thick,
inner sep=0pt,minimum size=1cm]
\tikzstyle{Dfive}=[circle, cross, draw=black!50,thick, inner sep=0pt,minimum size=0.4cm]
\tikzstyle{DfiveBig}=[circle, cross, draw=black!50,thick, inner sep=0pt,minimum size=0.6cm]
\tikzstyle{lpre}=[-,shorten <=0pt,>=stealth', very thick]
\tikzstyle{lpost}=[-,shorten >=0pt,>=stealth',very thick]
\tikzstyle{global}=[rectangle,draw=black!50,fill=black!20,thick,
inner sep=0pt,minimum size=1cm]
\tikzstyle{pre}=[<-,shorten <=0pt,>=stealth', very thick]
\tikzstyle{post}=[->,shorten >=0pt,>=stealth',very thick]
\tikzstyle{bpost}=[->, shorten >=2pt, shorten <=2pt, >=stealth', very thick]
\tikzstyle{bpre}=[<-, shorten >=2pt, shorten <=2pt, >=stealth', very thick]
\tikzstyle{nodo}=[circle,draw=black,fill=black,thick, inner sep=0pt,minimum size=2mm]
\tikzstyle{nodoblu}=[circle,draw=blue,fill=blue,thick, inner sep=0pt,minimum size=2mm]
\tikzset{cross/.style={path picture={ 
  \draw[black]
(path picture bounding box.south east) -- (path picture bounding box.north west) (path picture bounding box.south west) -- (path picture bounding box.north east);
}}}
\newcommand{\cercle}[4]{
\node[circle,inner sep=0,minimum size={2*#2}](a) at (#1) {};
\draw[dashed, very thick] (a.#3) arc (#3:{#3+#4}:#2);
}
\tikzset{
    partial ellipse/.style args={#1:#2:#3}{
        insert path={+ (#1:#3) arc (#1:#2:#3)}
    }
}

\begin{document}
\pagestyle{plain}

\makeatletter
\@addtoreset{equation}{section}
\makeatother
\renewcommand{\theequation}{\thesection.\arabic{equation}}

\thispagestyle{empty}

\vspace*{-2cm} 
\begin{flushright}
IFT-UAM/CSIC-16-087
\end{flushright}

\vspace*{0.8cm} 
\begin{center}
 {\LARGE Hilbert Series and Mixed Branches  of  $T[SU(N)]$ theories.}\\

 \vspace*{1.5cm}
{Federico Carta$^1$\footnote[$\ast$]{La Caixa-Severo Ochoa Scholar, e-mail: {\tt federico.carta@csic.es}}, Hirotaka Hayashi$^{1,2}$\footnote[$\dagger$]{e-mail: {\tt h.hayashi@tokai.ac.jp}}}\\

 \vspace*{1.0cm} 
$^1$ {\it Departamento de F\'isica Te\'orica and Instituto de F\'{\i}sica Te\'orica UAM-CSIC, \\Universidad Aut\'onoma de Madrid, Cantoblanco, 28049 Madrid, Spain}\\

\vspace*{0.5cm}
$^2$ {\it Tokai University, 4-1-1 Kitakaname, Hiratsuka, Kanagawa 259-1292, Japan }\\
  
\vspace*{0.8cm}
\end{center}
\vspace*{.5cm}

\noindent

We consider mixed branches of $3d$ $\mathcal{N}=4$ $T[SU(N)]$ theory. We compute the Hilbert series of the Coulomb branch part of the mixed branch from a restriction rule acting on the Hilbert series of the full Coulomb branch that will truncate the magnetic charge summation only to the subset of BPS dressed monopole operators that arise in the Coulomb branch sublocus where the mixed branch stems. This restriction can be understood directly from the type IIB brane picture by a relation between the magnetic charges of the monopoles and brane position moduli. We also apply the restriction rule to the Higgs branch part of a given mixed branch by exploiting 3d mirror symmetry. Both cases show complete agreement with the results calculated by different methods.

\newpage
\setcounter{page}{1}
\pagestyle{plain}
\renewcommand{\thefootnote}{\arabic{footnote}}
\setcounter{footnote}{0}


\tableofcontents


\section{Introduction}

One of the important aspects of supersymmetric quantum field theory are the geometry and structure of the moduli space of vacua and the associated chiral ring, which often provides insight on strongly-coupled regimes (for general reviews, see \cite{Intriligator:1995au, Strassler:2003qg}). Generically, the classical moduli space may receive quantum corrections and therefore the quantum moduli space can be quite different from the classical one. Hence, it is typically difficult to determine the quantum moduli space exactly. For example, three-dimensional (3d) $\mathcal{N}=4$ supersymmetric field theories have a Coulomb branch and a Higgs branch, which are exchanged with each other by 3d $\mathcal{N}=4$ mirror symmetry \cite{Intriligator:1996ex, deBoer:1996mp, Hanany:1996ie, deBoer:1996ck}. Although the Higgs branch is classically exact, the Coulomb branch moduli space is not protected against quantum corrections. Those theories are strongly-coupled at low energies, and the understanding of the exact Coulomb branch moduli space of 3d $\mathcal{N}=4$ supersymmetric field theories usually gives insight into strongly-coupled dynamics of the theories. 

Recently there has been much progress in the systematic determination of the exact Coulomb branch moduli space of 3d $\mathcal{N}=4$ supersymmetric gauge theories by using the so-called {\it Hilbert series approach}, which was initiated in \cite{Cremonesi:2013lqa}. The Hilbert series (HS) is a generating function counting chiral operators, graded by the charges they carry under the global symmetry group of the theory. It contains the information of chiral operators and their relations, and hence we can reconstruct the moduli space from the Hilbert series\footnote{A different approach has taken in \cite{Nakajima:2015txa, Bullimore:2015lsa, Braverman:2016wma, Bullimore:2016hdc} to construct the Coulomb branch chiral ring and its quantization.}. As for the Coulomb branch of 3d $\mathcal{N}=4$ supersymmetric gauge theories, one needs to take into account monopole operators, which can be defined as a disorder operator in the infrared (IR) superconformal field theory \cite{tHooft:1977hy, Borokhov:2002ib, Borokhov:2002cg, Borokhov:2003yu, Borokhov:2004kx, Bashkirov:2010hj}. The Hilbert series computation becomes possible by adopting the dimension formula \cite{Borokhov:2002cg, Gaiotto:2008ak, Benna:2009xd, Bashkirov:2010kz, Cremonesi:2013lqa} of the monopole operators as well as the fact that there is a unique BPS bare monopole operator for every magnetic charge \cite{Borokhov:2002cg}. Then the Hilbert series can be expressed as a summation over all possible magnetic charges. This method has been applied to various 3d $\mathcal{N}=4$ theories, for example, in \cite{Dey:2014tka, Cremonesi:2014kwa, Cremonesi:2014vla, Cremonesi:2014xha, Cremonesi:2014uva, Mekareeya:2015bla, Hanany:2015hxa, Hanany:2016ezz}.

The Hilbert series approach has been successfully extended to the calculation of the moduli space of 3d $\mathcal{N}=2$ gauge theories in \cite{Hanany:2015via, Cremonesi:2015dja, Cremonesi:2016nbo}. In 3d $\mathcal{N}=2$ gauge theories, non-perturbative superpotentials \cite{Affleck:1982as} can be generated and lift the classical Coulomb branch \cite{deBoer:1997kr, Aharony:1997bx}. The quantum corrections restrict the values of the Coulomb branch moduli and the restriction in turn is related to the restriction of magnetic charges. Therefore, the Hilbert series of the 3d $\mathcal{N}=2$ gauge theories theories can be written by a summation of magnetic charges, which is restricted by the effects of superpotentials. 

In \cite{Cremonesi:2015dja} it was further pointed out that the restriction may be also applied to the computation of the Hilbert series of mixed branched of 3d $\mathcal{N}=4$ $T[SU(N)]$ theory and this is indeed the direction we take in this paper. Namely, we apply the restriction technique to the computation of the Hilbert series of mixed branch of the 3d $\mathcal{N}=4$ $T[SU(N)]$ theory and indeed find agreement with the results calculated from a different method. 

The $T[SU(N)]$ theory arises as the S-dual to a half-BPS boundary condition of an $\mathcal{N}=4$ super Yang-Mills theory \cite{Gaiotto:2008ak}. The $T[SU(N)]$ theory is also related to regular punctures of four-dimensional class $\mathcal{S}$ theories \cite{Chacaltana:2012zy, Yonekura:2013mya}. Mixed branch of the 3d $T[SU(N)]$ theory may be classified by a Young tableaux with $N$ boxes, or equivalently a partition $\rho$ of the integer $N$ as done in \cite{Gaiotto:2008ak, Chacaltana:2012zy, Xie:2014pua}. The full moduli space is then given by 
\be
\mathcal{M}_{T[SU(N)]} = \cup_{\rho} \; \mathcal{C}_{\rho} \times \mathcal{H}_{\rho}, \label{TSUN.mixed}
\ee
where $\mathcal{C}_{\rho}$ is the Coulomb branch factor and $\mathcal{H}_{\rho}$ is the Higgs branch factor\footnote{For the special case where the mixed branch is $\mathcal{C}_{\rho} \times \{0\}$, we call the Coulomb branch as the full Coulomb branch since the branch has maximal dimension. Here $\{0\}$ stands for an origin. Similarly, we call the full Higgs branch for the Higgs branch part of the case of $\{0\} \times \mathcal{H}_{\rho}$.}. 
The mixed branch structure of the 3d $T[SU(N)]$ can also play an important role to determine the mixed branch of the 4d class $\mathcal{S}$ theories \cite{Xie:2014pua}.

The main aim of this paper is to compute the Hilbert series of the mixed branch $\mathcal{C}_{\rho} \times \mathcal{H}_{\rho}$ in \eqref{TSUN.mixed} from the restriction technique. Originally, the Hilbert series of the Coulomb branch part $\mathcal{C}_{\rho}$ has been computed in \cite{Cremonesi:2014kwa, Cremonesi:2014vla} by making use of a gauge theory whose full Coulomb branch moduli space gives $\mathcal{C}_{\rho}$ and also in \cite{Hanany:2011db} by utilizing the 3d mirror symmetry. We here compute the Hilbert series of the Coulomb branch part $\mathcal{C}_{\rho}$ from the Hilbert series of the full Coulomb branch of the $T[SU(N)]$ theory, by restricting the summation of the magnetic charges in the latter calculation. In fact, we argue that the restriction rule can be obtained directly from the brane configuration realizing the mixed branch of the 3d $T[SU(N)]$ theory. Hence, our method does not use the information of the IR gauge theory but only uses the brane configuration as well as the Hilbert series of the full Coulomb branch. The Hilbert series of the Higgs branch factor $\mathcal{H}_{\rho}$ can be also calculated with the restriction rule by using the mirror symmetry relation $\mathcal{H}_{\rho} \simeq \mathcal{C}_{\rho^D}$ \cite{Gaiotto:2008ak, Chacaltana:2012zy, Yonekura:2013mya} where $\rho^D$ is dual to the partition $\rho$. We also give a way to compute the Hilbert series of the Higgs branch factor $\mathcal{H}_{\alpha}$ by applying the technique developed in \cite{Benvenuti:2006qr, Gray:2008yu} to a 3d $\mathcal{N}=4$ gauge theory whose full Higgs branch is isomorphic to $\mathcal{H}_{\rho}$, which yields a check of the restriction rule as well as the 3d mirror symmetry.

This paper is organized as follows. In section \ref{section2} we briefly review the main underlying idea to the ``Hilbert Series Program", as well as the procedure to compute explicitly the Hilbert series for the full Coulomb and the full Higgs branch of the moduli space of a generic Lagrangian $3d$ $\mathcal{N}=4$ supersymmetric gauge theory. 
In section \ref{section3} we focus on the mixed branch of the $T[SU(N)]$ theory. We present a method of computing the Hilbert series of the Coulomb branch and the Higgs branch parts of the mixed branch of the $T[SU(N)]$ theory by using the techniques reviewed in section \ref{section2}.  We then use in section \ref{sec:restriction} the restriction rule to compute the Hilbert series of the Coulomb branch part of a mixed branch of the $T[SU(N)]$ theory from the Hilbert series of the full Coulomb branch of the $T[SU(N)]$ theory. We describe how we can obtain the restriction rule from the brane picture realizing the mixed branch of the $T[SU(N)]$ theory. In section \ref{section5}, we give some examples of how this procedure works, and we perform some consistency checks.
In section \ref{sec:Higgs}, we also compare the Hilbert series of the Higgs branch moduli space computed by the method in section \ref{section2} with the one obtained by using the restriction rule in section \ref{sec:restriction} as well as the 3d mirror symmetry. We finally summarize our results in section \ref{sec:conclusion} and also give a speculative argumet for computing the Hilbert series of the full moduli space of the $T[SU(N)]$ theory by utilizing the restriction rule and then gluing the different mixed branches altogether.

\section{Hilbert Series for Moduli Spaces of 3d $\mathcal{N}=4$ Theories}
\label{section2}

In this section, we will briefly review the Hilbert Series technique for studying the moduli space of three-dimensional $\mathcal{N}=4$ supersymmetric gauge theories, which was first developed in \cite{Benvenuti:2006qr, Gray:2008yu} for the full Higgs branch and in \cite{Cremonesi:2013lqa} for the full Coulomb branch.
This method was successfully tested in different contexts and already produced some interesting applications, for example the computation of the moduli spaces of instantons in \cite{Benvenuti:2010pq, Hanany:2012dm,Dey:2013fea, Cremonesi:2014xha, Mekareeya:2015bla, Hanany:2015hxa}. Here we recall the minimal notions needed in the following, and we refer to the literature for more details.

For a generic field theory, the moduli space $\mathcal{M}$ is defined as a set of gauge-inequivalent vacua.
Each point of $\mathcal{M}$ is labeled by vacuum expectation values (vevs) of a set of scalar fields of the theory, which therefore give coordinates on $\mathcal{M}$. In a $3d$ $\mathcal{N}=4$ supersymmetric theory, the scalar fields belong either to the vector multiplets or to the hypermultiplets.
It is also known that for 3d $\mathcal{N}=4$ supersymmetric theories the geometry of $\mathcal{M}$ will be locally a product of a Higgs branch factor $\mathcal{H}$ and a Coulomb branch factor $\mathcal{C}$, where $\mathcal{H}$ is parameterized by the vev of the scalars in the hypermultiplets, and $\mathcal{C}$ is parameterized by the vev of the scalars in the vector multiplets. Moreover, both moduli spaces are HyperK$\ddot{\rm{a}}$hler varieties. 

In order to study the geometry of $\mathcal{M}$, a very fruitful approach is to count all the gauge-invariant chiral operators, grading them by their charges under all the different global symmetries of the theory.
In the following we briefly review why this is a good strategy to employ.

The chiral ring $\mathcal{R}$ of a supersymmetric field theory is defined as a ring of chiral operators. 
This ring is believed to be isomorphic to the ring $\mathcal{O}$ of holomorphic functions defined over $\mathcal{M}$. In particular, we can associate a holomorphic function on the moduli space $\mathcal{M}$ to every element in $\mathcal{R}$. 
Now, if the ring of holomorphic functions on an unknown algebraic variety $\mathcal{M}$ is known, one can reconstruct and define $\mathcal{M}$ via usual techniques in algebraic geometry.
Namely, $\mathcal{M}$ will be defined as a scheme locally isomorphic to the spectrum of the ring $\mathcal{R}$, with Zariski's topology. 
While in principle this strategy will work, it is in general hard to explicitly determine all the elements of the chiral ring (or equivalently all the holomorphic functions on the moduli space), so one settles down to a more modest approach of simply counting chiral operators, grading them by their charges under all the symmetries that the theory under study enjoys. This is a well defined problem which is in general much simpler than computing the chiral ring exaclty.

The Hilbert series $HS(t)$ is the main tool used for this counting purpose. It is a generating function that keeps track, in a systematic way, of all the operators of the chiral ring. In more details, the coefficient $a_n$ in the Taylor expansion
\begin{equation}
HS(t)=\sum_{n}^{}a_nt^n,
\end{equation} 
will be equal to the number of  chiral operators having charge $n$ under the symmetry which is weighted by a fugacity $t$. 
This can be refined to the case in which one whishes to grade the chiral operators by more than one symmetry. For example, suppose that the chiral operators are charged under $N$ global symmetries. For each one of them, we choose $x_i,\ i=1,\cdots, N$ as a grading parameter. Then the Hilbert series will be given by
\begin{equation}
HS(t,x_i)=\sum_{k_1}^{}\sum_{k_2}^{}\cdots\sum_{k_N}^{} a_{k_1,k_2,\cdots k_n} \prod_{i=1}^{N} x_i^{k_i},
\end{equation}
and the interpretation is that $a_{k_1,k_2,\cdots k_N}$ is the number of chiral operators having respectively charges $k_1,k_2,\cdots k_N$ under the $N$ symmetries.

The algorithmic procedure to compute the Hilbert series from the data defining $d=3$ $\mathcal{N}=4$ gauge theories varies, depending on the fact that we want to compute the Hilbert series for the Higgs branch factor $\mathcal{H}$ or the Coulomb branch factor $\mathcal{C}$. Therefore we will split the discussion in two.

\subsection{Coulomb branch moduli space}
\label{sec:HSCoulomb}

The (full) Coulomb branch $\mathcal{C}$ of a $3d$ $\mathcal{N}=4$ supersymmetric gauge theory is characterized by giving nonzero vev to the triplet of scalars in the vector multiplets, and also by the vev of the dual photons. On a generic vacuum of the Coulomb branch the gauge group $G$ is broken to the maximal torus $U(1)^r$ where $r$ is the rank of the gauge group, and all the W-bosons and charged matter fields get massive. The geometry of $\mathcal{C}$ is a HyperK$\ddot{\rm{a}}$hler variety of the quaternion dimension equal to the rank $r$ of $G$.

Unlike the Higgs branch, the Coulomb branch receives quantum corrections.
An approach to the study of this branch employs monopole operators \cite{tHooft:1977hy, Borokhov:2002ib, Borokhov:2002cg, Borokhov:2003yu, Borokhov:2004kx, Bashkirov:2010hj}, i.e. disorder operators, analogous to the $4d$ 't Hooft operators. A bare monopole operator $V_m(x)$ is defined as a boundary condition in the Euclidean path integral, by requiring that the set of gauge connections onto which the path integral is performed will be restricted to a set of connections having a Dirac monopole's singularity (specified by an embedding $U(1)\mapsto G$) at the insertion point $x$.
Namely
\begin{equation}
A_{\pm}\sim \dfrac{m}{2}(\pm 1-\cos\theta)d\varphi, \label{monopoleconfig}
\end{equation}
where spherical coordinates $(r,\theta, \varphi)$ are used and $A_{\pm}$ is the gauge connection on the northern (respectively southern) hemisphere of a sphere $S^2$ surrounding the insertion point $x$. Here $m$ is the magnetic charge of the monopole operator, which takes values in the weight lattice of the Langlands (GNO) dual group $^LG$ \cite{Goddard:1976qe}, and satisfies a Dirac quantization condition \cite{Englert:1976ng}
\begin{equation}
\exp\left(2\pi i m\right)=1_G.
\label{Dirac}
\end{equation}

A bare monopole operator carries a magnetic charge $m$, defined as the flux of the gauge field through a sphere surrounding the insertion point of the monopole operator. It also has a conformal dimension, determined by its IR $R$-charge. Then the conformal dimension of a BPS bare monopole operator is given in terms of the magnetic charge by the following dimension formula \cite{Borokhov:2002cg, Gaiotto:2008ak, Benna:2009xd, Bashkirov:2010kz, Cremonesi:2013lqa}\footnote{The dimension formula for a monopole operator will be valid when the UV $U(1)_R$ symmetry is euqal to the IR superconformal R-symmetry. Those theories are called ``good'' or ``ugly'' in \cite{Gaiotto:2008ak}. In this article, we focus on a particular good theory called $T[SU(N)]$ theory, which we will define later. }
\begin{align}
\Delta(m)=-\sum_{\alpha\in \Delta^{+}}\left|\alpha(m)\right|+\dfrac{1}{2}\sum_i\sum_{\rho_i}\left|\rho_i(m)\right|,
\label{DimensionFormula}
\end{align}
where $\alpha$ are the positive roots of the gauge algebra, and $\rho_i$ are the weights of the matter representations.

The relevant operators in the chiral ring of the full Coulomb branch are however not bare monopole operators, but dressed monopole operators. Indeed, it is also possible to turn on a vev for a complex scalar in the adjoint representation of the vector multiplet, without spoiling the BPS condition \cite{Borokhov:2003yu}.

In order to count those operators grading them by their conformal dimension, it is crucial that there is exactly one bare BPS monopole operator for every magnetic charge $m$ \cite{Borokhov:2002cg}. However, there are still different ways in which it can be dressed. Given this, the Hilbert series is defined as
\begin{align}
HS(t)=\sum_{m\in\Gamma(^LG)/\mathcal{W}_{^LG}}t^{\Delta(m)}P_{G}(m,t)
\end{align}
where $t$ is a fugacity keeping track of the conformal dimension of the monopoles. The magnetic charge $m$ runs over all the lattice points of a Weyl chamber, i.e. over the weight lattice $\Gamma(^LG)$ of the Langlands (GNO) dual group of the gauge group modded out by the action of the Weyl group $\mathcal{W}_{^LG}$ \cite{Kapustin:2005py} .
Now, $P_{G}(m,t)$ is a correction factor taking care of the different dressings.

In details, the factor $P_{G}(m,t)$ is included due to the following reason. When the vev of a bare monopole operator is turned on in the background, the gauge group is generically broken to a subgroup $H_m\subset G$, defined as the subgroup of $G$ which commutes with the magnetic flux with the magnetic charge $m$. Then one can consistently turn on a vev for a complex scalar in the adjoint representation of this residual gauge group $H_m$, without spoiling the BPS conditions for the monopole.
$P_G(t,m)$ counts the gauge invariant operators of the residual group $H_m$. The explicit expression is given by
\begin{align}
P_G(t,m)=\prod_{i=1}^{r}\dfrac{1}{1-t^{d_i(m)}},
\end{align}
where $r$ is the rank of $H_m$ and $d_i(m)$ are all the degrees of the $r$ Casimir operators of $H_m$.
As a reference, the degrees of the Casimir operators are given in table \ref{tb:Casimir}.
\begin{table}[t]
\centering
\begin{tabular}{|l|l|}
\hline 
Simple Lie Algebra $\mathfrak{g}$ & Degrees \\ 
\hline 
$a_l, \quad l\geq 1$ & $2,3,\cdots, l+1$ \\ 
\hline 
$b_l, \quad l\geq 2$ & $2,4,\cdots, 2l$ \\
\hline
$c_l, \quad l\geq 3$ & $2,4,\cdots, 2l$ \\
\hline
$d_l, \quad l\geq 4$ & $2,4,\cdots, 2l-2, l$ \\
\hline
$e_6 \quad$ & $2,5,6,8,9,12$ \\
\hline
$e_7 \quad$ & $2,6,8,10,12,14,18$ \\
\hline
$e_8 \quad$ & $2,8,12,14,18,20,24,30$ \\
\hline
$f_4 \quad$ & $2,6,8,12$ \\
\hline
$g_2 \quad$ & $2,6$ \\
\hline
\end{tabular} 
\caption{Degrees of the Casimir invariants of the simple Lie algebras.}
\label{tb:Casimir}
\end{table}

In the case in which the gauge group $G$ consists of a product $G=\prod_iG_i$ of factors, and some of them are not simply connected, one can further refine this counting by including fugacities $z_i$ which keep track of charges under the 3d topological $U(1)^n_J$ symmetry. The topological $U(1)_J$ symmetry is a symmetry which induces in the semiclassical picture the shift of the dual photon \cite{Polyakov:1976fu}. The Hilbert series with this latter fugacities included, called now \emph{Refined Hilbert Series}, is then given by
\begin{align}
HS(t)=\sum_{m\in\Gamma(^LG)/\mathcal{W}_{^LG}}t^{\Delta(m)}\prod_{i=1}^nz_i^{J_i(m)}P_{G}(m,t), \label{HSC}
\end{align}
where $J_i(m)$ represents the charge of the monopole operator under the $i$-th $U(1)_{J}$ topological symmetry, where here $i=1,\cdots n$, and $n$ is the number of non-simply connected factors of $G$.

\subsection{Higgs branch moduli space}
\label{sec:HSHiggs}
The (full) Higgs branch of the moduli space of a $3d$ $\mathcal{N}=4$ supersymmetric gauge theory is characterized by giving nonzero vev to the scalars in the hypermultiplets. On a generic vacuum of the Higgs branch the gauge group is completely broken.
Unlike the Coulomb branch, the Higgs branch is protected against quantum corrections, and therefore its exact geometry can be studied in the classical theory.

In the Higgs branch, the relevant operators of the chiral ring are gauge invariant operators composed of hypermultiplets, subject to F-term conditions. One can count them by the following three step procedure if a Lagrangian of the theory is known \cite{Benvenuti:2006qr, Gray:2008yu}:
\begin{enumerate}
\item Generating all the possible symmetric products of the scalars in the chiral multiplets.\\
To do this one computes 
\begin{equation}
\PE\left[\sum_{i=1}^{2N_h}\mbox{char}_{R_i}(w)\mbox{char}_{R'_i}(x)\tilde{t}\right],
\end{equation}
where $w$ (resp. $x$) is a collective notation for all the $\rk(G_g)$ (resp. $\rk(G_f)$) fugacities of the gauge (resp. flavor) group. Also $\mbox{char}_{R_i}(w)$ (resp. $\mbox{char}_{R'_i}(x)$) is the character of the gauge (resp. flavor)
representation $R_i$ (resp. $R'_i$) of the $i$-th chiral multiplet $X_i$, $\tilde{t}$ is defined as $\tilde{t} = t^{\frac{1}{2}}$ where $t$ is again a fugacity counting the conformal dimension of some operators. Note that a scalar in 3d has dimension $\frac{1}{2}$. We introduced $\tilde{t}$ to avoid the appearance of fractional powers in the expressions. The sum is done over all the set of the $\mathcal{N}=2$ chiral multiplets belonging to $\mathcal{N}=4$ hypermultiplets. $N_h$ is the number of hypermultiplets and $2$ in front of $N_h$ appears since a $\mathcal{N}=4$ hypermultiplet is made of two $\mathcal{N}=2$ chiral multiplets. 
Here PE is the Plethysitic exponential, a generating function for symmetrizations, defined for any function $f(x_1,\cdots, x_n)$ such that $f(0,\cdots 0)=0$ as
\begin{equation}
\PE[f(x_1, \cdots, x_n)]:=\exp\left(\sum_{k=1}^{\infty}\dfrac{f(x_1^k,\cdots x_n^k)}{k}\right).
\label{symmetrizations}
\end{equation}

\item The F-term prefactor.

In this second step, one has to take into account the fact that the symmetric products of scalars generated in the step above are not independent, but subject to a number $N_r$ of relations arising from the fact that the $F$-term conditions need to be satisfied by every vacuum of the Higgs branch. To enforce this fact in the counting procedure, one has to multiply equation (\ref{symmetrizations}) by a factor
\begin{equation}
\mbox{Pfc}(w,\tilde{t}):=\PE\left[\sum_{i=1}^{N_r} \mbox{char}_{R''_i}(w) \tilde{t}^{d_i} \right]^{-1},
\end{equation}
where $\mbox{char}_{R''_i}(w)$ is the character of the gauge representation $R''_i$ of the $i$-th relation, and $d_i$ is its degree in the conformal dimension: typically $d_i=2$. The F-term relations will not usually depend on flavour fugacities, due to the fact that the superpotential (in terms of the $\mathcal{N}=2$ notation) involves a trace on the flavor indices, and this trace always appears also in the F-term equations. One might think that the variation under a hypermultiplet may give rise to an F-term equation that has flavor indices. However, the F-term condition is automatically satisfied since we do not turn on Coulomb branch moduli. Characters and degrees of the classical relations can be extracted easily from the superpotential, we will show detailed examples of how to do this in section \ref{sec:Higgs}.

\item The Molien-Weyl projection, (see e.g. \cite{opac-b1103234}).\\
In order to count only the gauge invariant operators, and not all of the symmetric products, we need to project all the representations that the PE generates onto the gauge singlets. This is done by integrating the gauge fugacities over the whole gauge group.
This indeed works since from representation theory it holds that $\int d\mu_{G} \ \mbox{char}_{R_i}(w)\mbox{char}_{\bar{R}_j}(w)=\delta_{ij}$. This implies that only gauge singlets give non-zero contribution after the integration. 
Therefore, integrating the result of step 1 and 2 over the full gauge group will discard all the gauge-variant operators, and keep only the gauge invariant ones.

In conclusion the Hilbert series of a Higgs branch is given by\footnote{Here we have used the well known property of the Plethystic exponential that $\PE[f(t)+g(t)]=\PE[f(t)]\PE[g(t)]$, for any $f(t)$ and $g(t)$ such that $f(0)=g(0)=0$.}
\begin{equation}
HS(\tilde{t},z)=\int_G d\mu_G \ \mbox{Pfc}(w,\tilde{t}) \prod_i PE\left[\mbox{char}_{R_i}(w)\mbox{char}_{R'_i}(z)\tilde{t}\right],\label{HSHiggs.comp}
\end{equation}
where $\mu_G$ is the Haar measure of $G$, defined for any Lie group as (see e.g. \cite{opac-b1099654})
\begin{equation}
\int_G d\mu_{G}=\dfrac{1}{(2\pi i)^r}\oint_{|w|_1=1}\cdots\oint_{|w|_r=1}\dfrac{dw_1}{w_1}\cdots\dfrac{dw_r}{w_r}\prod_{\alpha\in\Delta^+}\left(1-\prod_{k=1}^rw_k^{\alpha_k}\right),
\end{equation}
where $\Delta^+$ is the set of positive roots of the Lie algebra of $G$.
\end{enumerate}

\section{Mixed Branches of the $T[SU(N)]$ Theory}
\label{section3}
So far we have focused on a full Coulomb branch and a full Higgs branch of 3d $\mathcal{N}=4$ theories. In general, 3d $\mathcal{N}=4$ theories have many mixed branches where we can turn on vevs for scalars both in vector multiplets and hypermultiplets. For example, at some special locus of a full Coulomb branch, we may turn on vevs for scalars in hypermultiplets and there open up some directions in a Higgs branch. Then, the full moduli space of a generic three-dimensional $\mathcal{N}=4$ theory in fact has the structure 
\begin{equation}
\bigcup_{\alpha}\mathcal{C}_{\alpha}\times\mathcal{H}_{\alpha}
\label{FullModuli}
\end{equation}
where $\alpha$ labels the different mixed branches, $\mathcal{C}_{\alpha}$ is the Coulomb branch factor and $\mathcal{H}_{\alpha}$ is the Higgs branch factor. 
Both $\mathcal{C}_{\alpha}$ and $\mathcal{H}_{\alpha}$ are Hyperk$\ddot{\rm{a}}$hler varieties, where $\mathcal{C}_{\alpha}$ is parametrized by the vev of scalars in the vector multiplets and the dual photon and $\mathcal{H}_{\alpha}$ by scalars in the hypermultiplets. The union in equation (\ref{FullModuli}) is clearly not a disjoint union, as in general different mixed branches intersect with one another. With this notation, a full Coulomb branch is $\mathcal{C}\times\{0\}$ and a full Higgs branch is $\{0\}\times\mathcal{H}$. Those two full branches intersect at a single point, where typically the theory is a superconformal field theory.

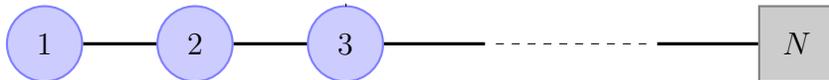
\begin{figure}[t]
\begin{center}
\begin{tikzpicture}
\node (m1) at ( -2,0) [gauge] {$1$};
\node (m2) at ( 0,0) [gauge] {$2$}
edge [lpre] node[auto,swap] {} (m1);
\node (stop) at (4,0) {}
edge [lpre] node[auto,swap] {} (m2);
\node (m3) at (2,0) [gauge] {$3$}
edge [lpre] node[auto,swap] {} (m3);
\node (stop2) at (6,0) {};
\draw [dashed] (4,0)--(6,0);
\node (n) at ( 8,0) [global] {$N$}
edge [lpre] node[auto,swap] {} (stop2);
\end{tikzpicture}
\caption{The quiver graph for the $T[SU(N)]$ theory. Here each circle node with a number $k, \  k=1, \cdots N-1$ denotes a factor $U(k)$ of the gauge group, and a line between two gauge nodes stand for one hypermultiplet in the bi-fundamental representation of the two gauge groups. The rightmost node with a number $N$ denotes a $SU(N)$ flavor group. In other words, there are $N$ hypermultiplets in the fundamental representation of $U(N-1)$ gauge group.
}
\label{quiver}
\end{center}
\end{figure}

In this paper, we are mainly interested in the mixed branches of the $T[SU(N)]$ theory, which is realized by the linear quiver of figure \ref{quiver}.
In order to visualize the mixed branch structure of the $T[SU(N)]$ theory, it is useful to engineer it with a brane construction in type IIB superstring theory and we will heavily make use of it.

\subsection{The brane realization of the mixed branch}

The type IIB brane system yielding 3d $\mathcal{N}=4$ supersymmetric gauge theories was first analyzed in \cite{Hanany:1996ie}. The configuration of the branes in the ten-dimensional spacetime is shown in table \ref{branepositions}. 
\begin{table}[t]
\begin{center}
\begin{tabular}{|c|c|c|c|c|c|c|c|c|c|c|}
\hline 
• & $x_0$ & $x_1$ & $x_2$ & $x_3$ & $x_4$ & $x_5$ & $x_6$ & $x_7$ & $x_8$ &$x_9$ \\ 
\hline 
D3 & - & - & - & x & x & x & - & x & x & x \\ 
\hline 
D5 & - & - & - & - & - & - & x & x & x & x \\ 
\hline 
NS5 & - & - & - & x & x & x & x & - & - & - \\ 
\hline 
\end{tabular}
\caption{The brane system realizing a 3d $\mathcal{N}=4$ theory. In this table ``x" means that the brane is pointlike in that direction, while ``-" means that it is extended in that direction.\label{branepositions}}
\end{center}
\end{table} 
In this configuration, some D3-branes are suspended between NS5-branes, and are of finite length in the $x^6$-direction.
Therefore, the worldvolume theory on the D3-branes is effectively a 3d $\mathcal{N}=4$ theory after the dimensional reduction along the $x_6$-direction.  
The rotational symmetry in the $(x_3, x_4, x_5)$-plane and in the $(x_7, x_8, x_9)$-plane gives the $SO(3) \times SO(3)$ R-symmetry of the 3d $\mathcal{N}=4$ supersymmetric theory.

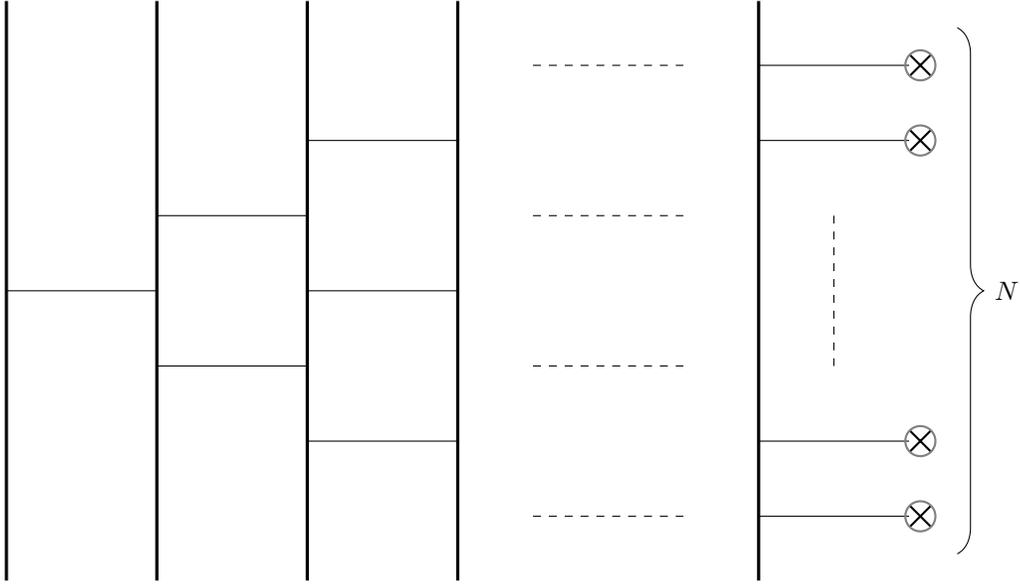
\begin{figure}[t]
\begin{center}
\begin{tikzpicture}
\node (u1u) at ( 0,4)  {};
\node (u1d) at ( 0,-4)  {}
edge [lpre] node[auto,swap] {} (u1u);
\node (u2u) at ( 2,4)  {};
\node (u2d) at ( 2,-4)  {}
edge [lpre] node[auto,swap] {} (u2u);
\node (u3u) at ( 4,4)  {};
\node (u3d) at ( 4,-4)  {}
edge [lpre] node[auto,swap] {} (u3u);
\node (u4u) at ( 6,4)  {};
\node (u4d) at ( 6,-4)  {}
edge [lpre] node[auto,swap] {} (u4u);
\node (uFu) at ( 10,4)  {};
\node (uFd) at ( 10,-4)  {}
edge [lpre] node[auto,swap] {} (uFu);
\draw (0,0)--(2,0);
\draw (2,1)--(4,1);
\draw (2,-1)--(4,-1);
\draw (4,2)--(6,2);
\draw (4,0)--(6,0);
\draw (4,-2)--(6,-2);
\draw [dashed] (7,3)--(9,3);
\draw [dashed] (7,1)--(9,1);
\draw [dashed] (7,-1)--(9,-1);
\draw [dashed] (7,-3)--(9,-3);

\draw (10,-2)--(12,-2);
\draw (10,2)--(12,2);
\draw (10,-3)--(12,-3);
\draw (10,3)--(12,3);
\draw [dashed] (11,1)--(11,-1);

\node [below] at (1,0.5) {};
\node [below] at (3,1.5) {};
\node [below] at (3,-0.5) {};
\node [below] at (5,2.5) {};
\node [below] at (5,0.5) {};
\node [below] at (5,-1.5) {};

\node [Dfive] (d51) at (12.15,3) {};
\node [Dfive] (d51) at (12.15,2) {};
\node [Dfive] (d52) at (12.15,-2) {};
\node [Dfive] (d53) at (12.15,-3) {};

\draw [decorate,decoration={brace,amplitude=10pt,mirror,raise=4pt},yshift=0pt]
(12.5,-3.5) -- (12.5,3.5) node [black,midway,xshift=0.8cm] {\footnotesize $N$};

\end{tikzpicture}
\caption{The brane picture for the $T[SU(N)]$ theory. In this picture the directions $x_0,x_1,x_2$ are suppressed, since they are shared among all the branes. The horizontal axis is $x_6$, the vertical axis corresponds to directions $x_7,x_8,x_9$ on which the NS5-branes are stretched, and the ``out of the page axis" corresponds to $x_3,x_4,x_5$, on which the flavor D5-branes are stretched. Hence, horizontal lines and vertical lines represent D3-branes and NS5-branes respectively. D5-branes are denoted by $\otimes$.}
\label{TSUN}
\end{center}
\end{figure}

Also the $T[SU(N)]$ theory can be realized by using a brane system in type IIB string theory, which arises as the S-dual of a half-BPS boundary condition of a 4d $\mathcal{N}=4$ super Yang-Mills theory \cite{Gaiotto:2008ak}.  
The brane configuration which yields the 3d $T[SU(N)]$ theory is given in figure \ref{TSUN}.
We have $k$ D3-branes between the $k$-th and the $(k+1)$-th NS5-brane\footnote{The order is counted from left to right. Namely, the leftmost NS5-branes is the first NS5-brane.} 
for $k=1, \cdots, N-1$, and $N$ D3-branes are attached only to the last NS5-brane. At the end of each rightmost D3-brane, we may put one D5-brane. The introduction of the D5-branes will be useful for reading off the Higgs branch. The $k$ D3-branes between NS5-branes give rise to a gauge group $U(k)$, and we call them ``color D3-branes". On the other hand, the $N$ D3-branes attached to the rightmost NS5-brane realize the $SU(N)$ flavor symmetry, and we call them ``flavor D3-branes".

While the $N$ D5-branes in the brane configuration for the $T[SU(N)]$ theory yield the perturbative $SU(N)$ flavor symmetry, the $N$ NS5-branes in fact realize non-perturbative $SU(N)$ global symmetry \cite{Gaiotto:2008ak}. From the quiver description of the $T[SU(N)]$ theory, we know that at least we have the $U(1)_J^{N-1}$ topological global symmetry. The $U(1)_J^{N-1}$ topological global symmetry is in fact enhanced to $SU(N)$ by the effect of monopole operators. Moreover, the $T[SU(N)]$ theory is self-mirror and the full Coulomb branch moduli space is isomorphic to the full Higgs branch moduli space. 

One nice feature about the brane picture is that the Coulomb branch moduli space, the Higgs branch moduli space and all the mixed branches can be pictorically understood from brane motions. The D3-branes suspended between NS5-branes can move along the NS5-branes. These degrees of freedom correspond to the Coulomb branch moduli of the 3d gauge theory\footnote{One the other hand, the positions of the flavor D3-branes in the $(x_7, x_8, x_9)$-directions are related to the mass parameters of the fundamental hypermultiplets. }.
When we tune the positions of the color D3-branes in the $(x_7, x_8, x_9)$-directions, the flavor D3-branes may be fractionated between D5-branes and can move between the D5-branes in the $(x_3, x_4, x_5)$-directions. These latter degrees of freedom correspond to the moduli parametrizing the Higgs branch. In particular, when all the positions of the color D3-branes are tuned to zero, the full Higgs branch opens up. Due to this construction, the non-perturbative $SU(N)$ global symmetry is associated to the Coulomb branch and the perturbative $SU(N)$ flavor symmetry is associated to the Higgs branch.
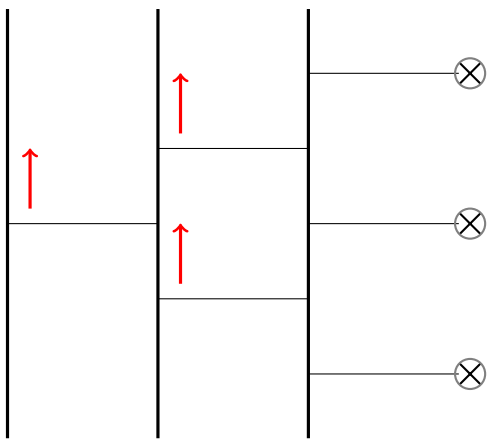
\begin{figure}[t]
\begin{center}
\begin{tikzpicture}
\node (u1u) at ( 0,3)  {};
\node (u1d) at ( 0,-3)  {}
edge [lpre] node[auto,swap] {} (u1u);
\node (u2u) at ( 2,3)  {};
\node (u2d) at ( 2,-3)  {}
edge [lpre] node[auto,swap] {} (u2u);
\node (u3u) at ( 4,3)  {};
\node (u3d) at ( 4,-3)  {}
edge [lpre] node[auto,swap] {} (u3u);
\draw (0,0)--(2,0);
\draw (2,1)--(4,1);
\draw (2,-1)--(4,-1);
\draw (4,2)--(6,2);
\draw (4,0)--(6,0);
\draw (4,-2)--(6,-2);
\node [Dfive] (d51) at (6.15,2) {};
\node [Dfive] (d52) at (6.15,0) {};
\node [Dfive] (d53) at (6.15,-2) {};
\draw [->, very thick, red] (0.3,0.2)--(0.3,1);
\draw [->, very thick, red] (2.3,-0.8)--(2.3,0);
\draw [->, very thick, red] (2.3,1.2)--(2.3,2);
\end{tikzpicture}
\caption{The brane picture for the full Coulomb branch of $T[SU(3)]$.}
\label{FullCoulombTSU3}
\end{center}
\end{figure} 
\begin{figure}[t]
\begin{center}
\begin{tikzpicture}
\node (u1u) at ( 0,3)  {};
\node (u1d) at ( 0,-3)  {}
edge [lpre] node[auto,swap] {} (u1u);
\node (u2u) at ( 2,3)  {};
\node (u2d) at ( 2,-3)  {}
edge [lpre] node[auto,swap] {} (u2u);
\node (u3u) at ( 4,3)  {};
\node (u3d) at ( 4,-3)  {}
edge [lpre] node[auto,swap] {} (u3u);
\draw (0,0)--(6,0);
\draw (2,-0.1)--(6,-0.1);
\draw (4,-0.2)--(6,-0.2);
\draw [dashed] (5.15,-1.1)--(7.15,0.9);
\draw [dashed] (7.15,-1.1)--(9.15,0.9);
\draw [dashed] (9.15,-1.1)--(11.15,0.9);
\draw (6.95,0.7)--(8.95,0.7);
\draw (8.35,-0.1)--(9.95,-0.1);
\draw (5.25,-1)--(7.25,-1);
\node [Dfive] (d51) at (6.15,-0.1) {};
\node [Dfive] (d52) at (8.15,-0.1) {};
\node [Dfive] (d53) at (10.15,-0.1) {};
\draw [->, very thick, blue] (7.2,0.8)--(7.7,1.3);
\draw [->, very thick, blue] (6,-0.9)--(6.5,-0.4);
\draw [->, very thick, blue] (9,0)--(9.5,0.5);
\end{tikzpicture}
\caption{The brane picture for the full Higgs branch of $T[SU(3)]$.}
\label{FullHiggsTSU3}
\end{center}
\end{figure}
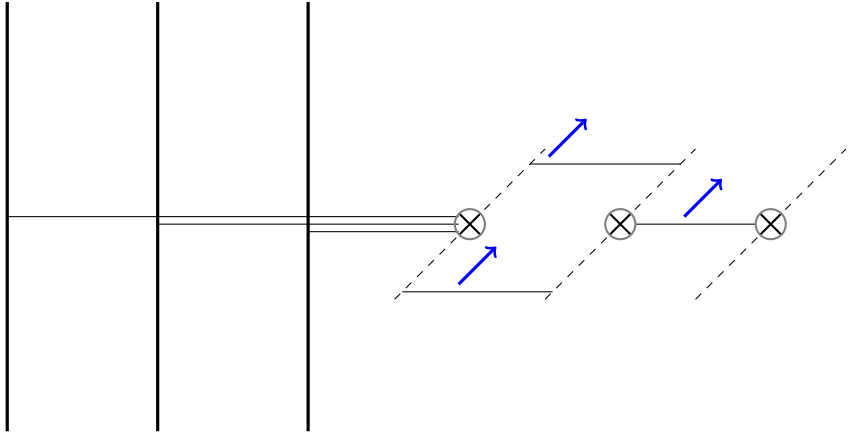
The full Coulomb branch and the full Higgs branch of the $T[SU(3)]$ theory is shown in figure \ref{FullCoulombTSU3} and \ref{FullHiggsTSU3} respectively.

A mixed branch of the $T[SU(N)]$ theory may arise when only a part of the positions of the color D3-branes are tuned. At some subloci of the full Coulomb branch moduli space, a Higgs branch opens up. In fact, the subloci where a Higgs branch opens up are given by nilpotent orbits of $\mathfrak{su}(N)$, and can be classified by a Young diagram with $N$ boxes or equivalently a partition of the integer $N$ \cite{Gaiotto:2008ak, Chacaltana:2012zy, Xie:2014pua}. The correspondence goes as follows. A partition $\rho = [a_1, a_2, \cdots, a_n]$ with $a_1 \geq a_2 \geq \cdots \geq a_n$ and $\sum_{i=1}^na_i = N$\footnote{In terms of a Young diagram, $\rho = [a_1, a_2, \cdots, a_n]$ means that the Young diagram has $a_i$ boxes for the $i$-th column for $i=1, \cdots, n$. Due to this correspondence, we will use a partition and the corresponding Young diagram interchangeably and write the Young diagram associated to a partition $\rho$ as $Y_{\rho}$.} means that $a_i$ flavor D3-branes are put together on one D5-brane for each $i=1, \cdots, n$. Note that this restriction does not only fix the positions of flavor D3-branes but also fix the positions of color D3-branes. This is due to the s-rule which states that only one D3-brane can be suspended between an NS5-brane and a D5-brane in order to preserve the supersymmetry \cite{Hanany:1996ie}. Therefore, when some flavor D3-branes are put on one D5-brane, some of the flavor D3-branes should connect to some color D3-branes so that the configuration does not break the s-rule. In this way, the Young diagram classification can tune the Coulomb branch moduli.

\begin{figure}[t]
\begin{center}
\begin{tikzpicture}
\node (u1u) at ( 0,3)  {};
\node (u1d) at ( 0,-3)  {}
edge [lpre] node[auto,swap] {} (u1u);
\node (u2u) at ( 2,3)  {};
\node (u2d) at ( 2,-3)  {}
edge [lpre] node[auto,swap] {} (u2u);
\node (u3u) at ( 4,3)  {};
\node (u3d) at ( 4,-3)  {}
edge [lpre] node[auto,swap] {} (u3u);
\draw (0,0)--(2,0);
\draw (2,1.9)--(4,1.9);
\draw (2,-1)--(4,-1);
\draw (4,2)--(6,2);
\draw (4,1.8)--(8,1.8);
\draw (8.55,2.3)--(10.55,2.3);
\draw (4,1.9)--(6,1.9);
\draw [dashed] (5.15,0.9)--(7.15,2.9);
\draw [dashed] (7.15,0.9)--(9.15,2.9);
\draw [dashed] (9.15,0.9)--(11.15,2.9);
\draw (6.95,2.7)--(8.95,2.7);
\node [below] at (1,0.5) {};
\node [below] at (3,2.5) {};
\node [below] at (3,-0.5) {};
\node [Dfive] (d51) at (6.15,1.9) {};
\node [Dfive] (d52) at (8.15,1.9) {};
\node [Dfive] (d53) at (10.15,1.9) {};
\draw [->, very thick, blue] (7.2,2.8)--(7.7,3.3);
\draw [->, very thick, red] (0.3,0.2)--(0.3,1);
\draw [->, very thick, red] (2.3,-0.8)--(2.3,0);
\draw [->, very thick, blue] (9,2.5)--(9.5,3);
\end{tikzpicture}
\caption{The brane picture for the mixed branch $\rho=[2,1]$.}
\label{CoulHiggs}
\end{center}
\end{figure}
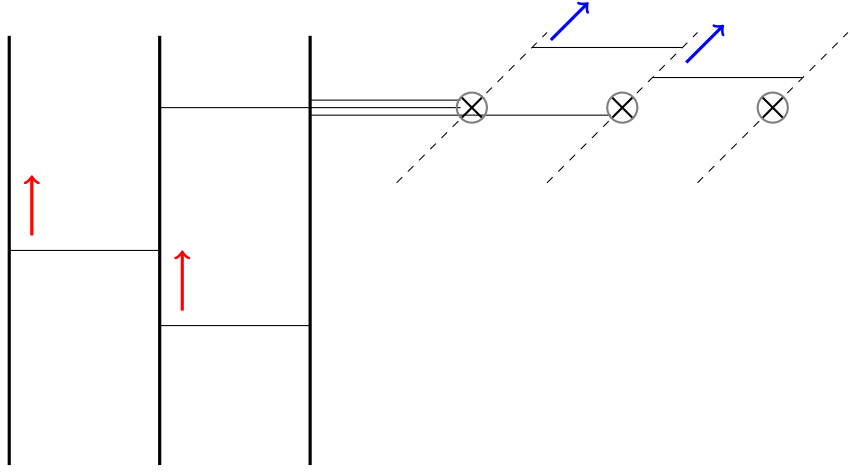

When some of the positions of the color D3-branes are fixed, some of the flavor D3-branes may be fractionated between D5-branes and hence a mixed branch of the $T[SU(N)]$ theory can be realized. Note that in order to realize the maximal Higgs branch of a mixed branch, one also needs to tune the mass parameters of the remaining fundamental hypermultiplets, An example of the mixed branch corresponding to the partition $\rho=[2, 1]$ of the $T[SU(3)]$ theory is shown in figure \ref{CoulHiggs}.

Since for the $T[SU(N)]$ theory the mixed branch structure may be completely specified by the partition $\rho$ with $N$ boxes \cite{Gaiotto:2008ak, Chacaltana:2012zy, Xie:2014pua}, the full moduli space is given by 
\begin{equation}
\bigcup_{\rho}\mathcal{C}_{\rho} \times \mathcal{H}_{\rho}, \label{mixedTSUN}
\end{equation}
where $\rho$ is all the possible partitions of the integer $N$. In particular, $\rho = [1, 1, \cdots, 1]$ gives $\mathcal{C} \times \{0\}$ with the maximal Coulomb branch $\mathcal{C}$, and $\rho = [N]$ gives $\{0\} \times \mathcal{H}$ with the maximal Higgs branch $\mathcal{H}$. The dimension of the Coulomb branch moduli space $\mathcal{C}_{\rho}$ can be computed from the associated partition \cite{Gaiotto:2008ak, Chacaltana:2012zy, Xie:2014pua}, 
\begin{equation}
\dim_{\mathbb{H}}(\mathcal{C}_{\rho})=\dfrac{1}{2}\left(N^2-\sum_{i=1}^{n}a_i^2\right),
\end{equation}
where $a_i, i=1, \cdots, n$ is the entry of the partition $\rho = [a_1, \cdots, a_n]$. For example, one can check
\begin{equation}
\dim_{\mathbb{H}}(\mathcal{C}_{[2,1]}) = \frac{1}{2}\left(3^2- (2^2+1^2)\right) = 2, \label{dim21}
\end{equation}
which agrees with the number of color D3-branes that are not frozen in figure \ref{CoulHiggs}.

In fact, the mirror symmetry of the 3d $\mathcal{N}=4$ theory implies \cite{Gaiotto:2008ak, Chacaltana:2012zy, Xie:2014pua}
\begin{equation}
\mathcal{H}_{\rho} \simeq \mathcal{C}_{\rho^D}, \label{Higgsdual}
\end{equation}
where $\rho^D$ is the dual partition to $\rho$, which is associated to the transpose of the Young diagram $Y_{\rho}$. This property can be inferred from the brane configuration. In terms of the brane configuration, the mirror symmetry is realized by the S-duality in type IIB string theory \cite{Hanany:1996ie}, which exchanges NS5-branes with D5-branes but keep D3-branes unchanged. Since the $T[SU(N)]$ theory is self-mirror, a Higgs branch $\mathcal{H}_{\rho}$ in a mixed branch specified by $\rho$ of the $T[SU(N)]$ theory is mapped to a Coulomb branch $\mathcal{C}_{\rho'}$ in a different mixed branch specified by a different partition $\rho'$ of the $T[SU(N)]$ theory . The partition $\rho'$ should be related to the number of flavor D3-branes put on one D5-brane in the mirror picture. Hence, in the original theory, $\rho'$ should be related to the number of D3-branes put on one NS5-brane. Suppose $\rho$ is given by $[a_1, \cdots, a_n]$ with $a_1 \geq a_2 \geq \cdots \geq a_n$ and $\sum_{i=1}^n a_i = N$. This means that for example $n$ D3-branes end on the rightmost NS5-brane. In general, if the number of $a_i$ satisfying $a_i \geq k$ is $b_k$, then there are $b_k$ D3-branes ending on the $(N-k+1)$-th NS5-brane. Therefore, we find that $\rho'$ is given by the partition $[b_1, \cdots, b_{n'}]$ where $b_k$ is the number of $a_i$ satisfying $a_i \geq k$ for $i=1, \cdots, n$. Then it is possible to see that the partition $\rho'$ defined in this way is nothing but the dual partition $\rho^D$, yielding the claim \eqref{Higgsdual}. 

Due to this feature, one can write the full moduli space \eqref{mixedTSUN} as
\begin{equation}
\bigcup_{\rho}\mathcal{C}_{\rho} \times \mathcal{C}_{\rho^D}, \label{mixedTSUN1}
\end{equation}
or
\begin{equation}
\bigcup_{\rho}\mathcal{H}_{\rho^D} \times \mathcal{H}_{\rho}. \label{mixedTSUN2}
\end{equation}
The relation \eqref{Higgsdual} also implies that the dimension of the Higgs branch $\mathcal{H}_{\rho}$ of the mixed branch specified by $\rho$ may be given by
\begin{equation}
\dim_{\mathbb{H}}(\mathcal{H}_{\rho})=\dfrac{1}{2}\left(N^2-\sum_{i=1}^{n'}b_i^2\right),
\end{equation}
where $b_i, i=1, \cdots, n'$ is the entry of the partition $\rho' = [b_1, \cdots, b_{n'}]$ which is dual to $\rho$. For example regarding the Higgs branch factor $\mathcal{H}_{[2, 1]}$ the dimension can be counted by using the dual partition which is the same as $[2, 1]$. Then the dimension of $\mathcal{H}_{[2, 1]}$ is again $2$ from \eqref{dim21}, which agrees with the number of mobile D3-branes suspended between D5-branes in figure \ref{CoulHiggs}.

\subsection{Hilbert series for the Coulomb branch factor}
\label{sec:HSCoulomb}
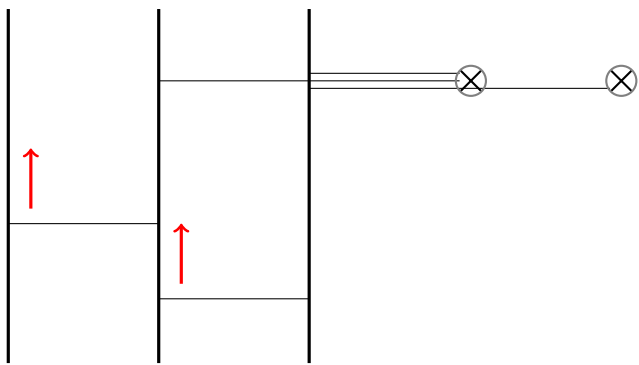
\begin{figure}[t]
\begin{center}
\begin{tikzpicture}
\node (u1u) at ( 0,3)  {};
\node (u1d) at ( 0,-2)  {}
edge [lpre] node[auto,swap] {} (u1u);
\node (u2u) at ( 2,3)  {};
\node (u2d) at ( 2,-2)  {}
edge [lpre] node[auto,swap] {} (u2u);
\node (u3u) at ( 4,3)  {};
\node (u3d) at ( 4,-2)  {}
edge [lpre] node[auto,swap] {} (u3u);
\draw (0,0)--(2,0);
\draw (2,1.9)--(4,1.9);
\draw (2,-1)--(4,-1);
\draw (4,2)--(6,2);
\draw (4,1.8)--(8,1.8);
\draw (4,1.9)--(6,1.9);
\node [below] at (1,0.5) {};
\node [below] at (3,2.5) {};
\node [below] at (3,-0.5) {};
\node [Dfive] (d51) at (6.15,1.9) {};
\node [Dfive] (d52) at (8.15,1.9) {};
\draw [->, very thick, red] (0.3,0.2)--(0.3,1);
\draw [->, very thick, red] (2.3,-0.8)--(2.3,0);
\end{tikzpicture}
\caption{The brane picture of the $T^{[2, 1]}[SU(3)]$ theory.}
\label{IRof21}
\end{center}
\end{figure}

It is possible to compute the Hilbert series for the Coulomb branch factor $\mathcal{C}_{\rho}$ in a mixed branch specified by $\rho$ by utilizing the method described in section \ref{section2}. Since the mixed branch is locally given by a product of the Coulomb branch factor $\mathcal{C}_{\rho}$ and the Higgs branch factor $\mathcal{H}_{\rho}$, the value of the vevs parameterizing $\mathcal{H}_{\rho}$ does not affect the Coulomb branch part $\mathcal{C}_{\rho}$. Hence, in particular we can consider infinitely large vevs for the scalars parameterizing $\mathcal{H}_{\rho}$. In terms of the brane picture, we send the pieces of D3-branes between D5-branes to infinity. At low energies at the infinitely large vev of the Higgs branch $\mathcal{H}_{\rho}$, one obtains a different 3d $\mathcal{N}=4$ theory which we call $T^{\rho}[SU(N)]$ theory. An example of the brane picture realizing the $T^{[2,1]}[SU(3)]$ theory is shown in figure \ref{IRof21}.

Since the Coulomb branch moduli space of the $T^{\rho}[SU(N)]$ theory should be the same as $\mathcal{C}_{\rho}$, one can consider the Hilbert series of the Coulomb branch for the $T^{\rho}[SU(N)]$ theory. The Hilbert series can be calculated by going to the gauge theory description of the $T^{\rho}[SU(N)]$ theory \cite{Cremonesi:2014kwa, Cremonesi:2014vla}. Although it is non-trivial to read off the gauge theory content from the original brane picture with several D3-branes on top of one D5-brane, one can move the D5-brane to the left until no D3-branes are attached to the D5-brane. The annihilation of D3-branes is due to the Hanany-Witten transitions. Then the D5-brane gives a hypermultiplet in the fundamental representation under the gauge group given by color D3-branes in the cell where the D5-brane is located. Once we obtain the gauge theory description of the $T^{\rho}[SU(N)]$ theory, we can use the method described in section \ref{section2} to compute the Hilbert series of the Coulomb branch of the $T^{\rho}[SU(N)]$ theory, which should coincide with the Hilbert series for $\mathcal{C}_{\rho}$. 

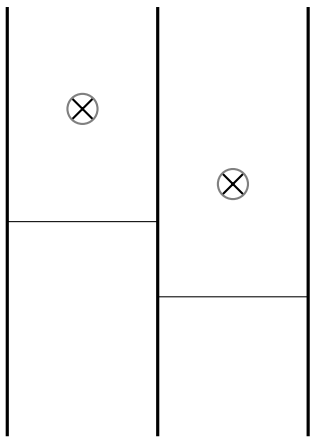
\begin{figure}[t]
\begin{center}
\begin{tikzpicture}
\node (u1u) at ( 0,3)  {};
\node (u1d) at ( 0,-3)  {}
edge [lpre] node[auto,swap] {} (u1u);
\node (u2u) at ( 2,3)  {};
\node (u2d) at ( 2,-3)  {}
edge [lpre] node[auto,swap] {} (u2u);
\node (u3u) at ( 4,3)  {};
\node (u3d) at ( 4,-3)  {}
edge [lpre] node[auto,swap] {} (u3u);
\draw (0,0)--(2,0);
\draw (2,-1)--(4,-1);
\node [Dfive] (d51) at (1,1.5) {};
\node [Dfive] (d51) at (3,0.5) {};
\end{tikzpicture}
\caption{The brane picture for the IR theory $T^{[2,1]}[SU(3)]$ after Hanany-Witten transitions compared with the one in figure \ref{IRof21}.}
\label{IR1111}
\end{center}
\end{figure}

The brane picture of the $T^{[2,1]}[SU(3)]$ case after the Hanany-Witten transitions  is given in figure \ref{IR1111}. To read off the gauge theory content we moved the two D5-branes in figure \ref{IRof21} to the left and obtain antoher brane configuration in figure \ref{IR1111}. From the brane configuration in figure \ref{IR1111} the gauge theory description can be inferred as
\begin{equation}
[1] - U(1) - U(1) - [1].
\end{equation} 
Here $[1]-$ or $-[1]$ is one hypermultiplet charged under the $U(1)$ to which the line is connected. The other line between the two $U(1)$'s denotes a hypermultiplet in the bi-fundamental representation under the gauge group $U(1) \times U(1)$. Similarly, we will use a notation where $[n]-$ implies $n$ hypermultiplets in the fundamental representation of the gauge group to which the line is connected and a line between two gauge groups means a hypermultiplet in the bi-fundamental representation of the two gauge groups.

In general, the $T^{\rho}[SU(N)]$ theory where $\rho = [a_1, \cdots, a_n]$ with $a_1 \geq a_2 \geq \cdots \geq a_n$ and $\sum_{i=1}^n a_i = N$ is given by the following linear quiver theory,
\begin{eqnarray}
{\overset{\overset{\text{\normalsize$\left[\#\left(a_i = N-1\right)\right] $}}{~~~\textstyle\vert}}{ U\left(1-N_1\right)}}  - \cdots - {\overset{\overset{\text{\normalsize$\left[\#\left(a_i=N-k\right)\right]$}}{~~~\textstyle\vert}}{U\left(k-N_k\right)}} 
- \cdots-  {\overset{\overset{\text{\normalsize$\left[\#\left(a_i = 1\right)\right]$}}{~~~\textstyle\vert}}{U\left(N-1-N_{N-1}\right)}}, \label{generalquiver}
\end{eqnarray}
with
\begin{equation}
N_k = \sum_{i=1}^n \left(a_i-(N-k)\right)H\left(a_i-(N-k)\right),\quad  k=1, \cdots, N-1, \label{Nk}
\end{equation}
where $H(x)$ is the Heaviside step function with the convention $H(0) = 0$ and $\#(a_i = l)$ is the number of $a_i$ which is equal to $l$ for $i=1, \cdots, n$. 

In the next section, we will describe a different technique, namely the restriction prescription, to compute the Hilbert series of the Coulomb branch factor $\mathcal{C}_{\rho}$. The  method in fact directly uses the brane picture realizing the mixed branch specified by a partition $\rho$ and does not use the IR gauge theory of $T^{\rho}[SU(N)]$.

\subsection{Hilbert series for the Higgs branch factor}
\label{sec:HSHiggspart}
It is also possible to calculate the Hilbert series of the Higgs branch factor $\mathcal{H}_{\rho}$ of a mixed branch specified by a partition $\rho$ by utilizing the method for computing the full Higgs branch described in section \ref{section2}. We can again make use of the locally product structure of the mixed branch. Namely, the Higgs branch factor $\mathcal{H}_{\rho}$ is independent of the value of the Coulomb branch moduli of $\mathcal{C}_{\rho}$. In particular, we can take infinitely large vevs for the Coulomb branch moduli. In terms of the brane picture, we send the non-fixed positions of the color D3-branes to infinity. At low energies at the infinitely large vev of the Coulomb branch moduli, one obtains a different theory which we call $\tilde{T}^{\rho}[SU(N)]$ theory. The resulting brane configuration of the the $\tilde{T}^{\rho}[SU(N)]$ theory is the one at the origin of the Coulomb branch of the $\tilde{T}^{\rho}[SU(N)]$ theory. By moving to a generic point of the Coulomb branch moduli space, one can read off the gauge theory content of the $\tilde{T}^{\rho}[SU(N)]$ theory. After knowing the gauge theory description, one can apply the technique for computing the Hilbert series of the Higgs branch introduced in section \ref{section2} to the gauge theory corresponding to the $\tilde{T}^{\rho}[SU(N)]$ theory. The full Higgs branch of the $\tilde{T}^{\rho}[SU(N)]$ theory should be the same as the Higgs branch factor $\mathcal{H}_{\rho}$ of the mixed branch. Similarly, both the Hilbert series should be the same.

\begin{figure}[t]
\begin{center}
\begin{tikzpicture}
\node (u1u) at ( 0,3)  {};
\node (u1d) at ( 0,-3)  {}
edge [lpre] node[auto,swap] {} (u1u);
\node (u2u) at ( 2,3)  {};
\node (u2d) at ( 2,-3)  {}
edge [lpre] node[auto,swap] {} (u2u);
\draw (0,-1)--(2,-1);
\draw (2,0)--(4,0);
\draw (2,2)--(4,2);
\draw (2,-2)--(4,-2);
\node [Dfive] (d51) at (4,2) {};
\node [Dfive] (d51) at (4,0) {};
\node [Dfive] (d51) at (4,-2) {};
\draw [->, very thick, red] (0.3,-0.8)--(0.3,0);
\end{tikzpicture}
\caption{The brane picture for the IR theory of $\tilde{T}^{[2, 1]}[SU(3)]$ obtained by decoupling all the unfrozen Coulomb branch moduli of the UV theory.}
\label{fig:U13flavors}
\end{center}
\end{figure}
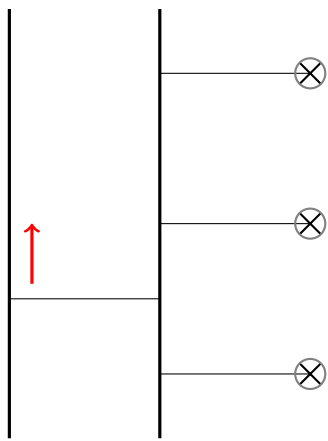

For example, as for the Higgs branch of the mixed branch specified by the partition $[2, 1]$ of the $T[SU(3)]$ theory, decoupling the Coulomb branch moduli yields the $U(1)$ gauge theory with $3$ flavors as in figure \ref{fig:U13flavors}.
Therefore, the Higgs branch factor $\mathcal{H}_{[2, 1]}$ is isomorphic to the full Higgs branch of the $U(1)$ gauge theory with $3$ flavors. 

In general, the IR theory at the infinitely large vev for the Coulomb branch part of the mixed branch specified by $\rho =[a_1, \cdots, a_n]$ with $a_1 \geq a_2 \geq \cdots \geq a_n$ and $\sum_{i=1}^n a_i = N$ is give by the following linear quiver theory,
\begin{equation}
U(N_1) - U(N_2) - \cdots - U(N_{N-1}) - [N], \label{IRHiggs}
\end{equation}
where $N_k, k=1, \cdots, N-1$ is given by \eqref{Nk}. When $N_k$ is zero then we remove the gauge node as well as the line attached to it.

\section{The Restriction Rule for the Hilbert Series}
\label{sec:restriction}

In this section we develop the main result of this article. We conjecture that the Hilbert series for the Coulomb branch part of a mixed branch can be obtained from the Hilbert series of the full Coulomb branch, by performing a specific restriction of the latter. We also explain how this restriction rule is easily understood in terms of the type IIB brane picture.

\subsection{The restriction rule}
\label{sec:rule}

In section \ref{sec:HSCoulomb}, we described a way to compute the Hilbert series of the Coulomb branch factor in the mixed branch specified by a partition $\rho$. For that, we made use of the gauge theory description of the $T^{\rho}[SU(N)]$ obtained after certain Hanany-Witten transitions of the corresponding brane diagram. However, we argue that we are able to compute the Hilbert series of the Coulomb branch factor without going to the gauge theory description but directly from the brane configuration realizing the mixed branch $\mathcal{M}_{\rho}$. 

Due to the boundary condition \eqref{monopoleconfig} at the insertion point of a monopole operator, the BPS condition implies that the real scalar $\sigma$ in the $\mathcal{N}=2$ vector multiplet inside the $\mathcal{N}=4$ vector multiplet satisfies \cite{Borokhov:2003yu}
\begin{equation}
\sigma \sim \frac{m}{2r} \label{BPS}
\end{equation}
where $m$ is the magnetic charge and $r$ is the radial coordinate. On the other hand, vevs of the scalars in the vector multiplet are related to color D3-brane positions. We can therefore relate, in the brane picture, color D3-brane positions with the magnetic charges of monopole operators. 

At a point in the Coulomb branch factor of a mixed branch, we tune the positions of some of the color D3-branes so that they coincide with the positions of flavor D3-branes ending on D5-brane. Since the positions of the color D3-branes are the Coulomb branch moduli and the positions of the flavor D3-branes are the mass parameters for fundamental hypermultiplets, the tuning implies that the Coulomb branch moduli are equal to the mass parameters. In order to obtain a mixed branch, we turn off the mass parameters and all the flavor D3-branes are aligned along one line. Then, it is possible to set the values of the masses to zero without loss of generality. This in turn means that the value of the frozen positions of the color D3-branes or equivalently the corresponding Coulomb branch moduli are zero. Then the BPS condition \eqref{BPS} means that the corresponding magnetic charges also have to be zero.

Hence, the restriction of the positions of the color D3-branes given by the partition $\rho$ can be translated into the condition that the corresponding magnetic charges are zero. Then, when one computes the Hilbert series of the Coulomb branch factor $\mathcal{C}_{\rho}$, one can simply insert the condition that some magnetic charges are zero into the Hilbert series for the full Coulomb branch. And the restriction of the magnetic charges can be read off from which color D3-branes are frozen. Physically, the restriction truncates the magnetic charges to a subset corresponding to BPS monopole operators that arise in the Coulomb branch factor.

In more detail, our conjecture of the Hilbert series of a Coulomb branch moduli space of a mixed branch specified by $\rho$ is 
\begin{align}
HS_{\rho}(t,z_i)=\sum_{\left.m_1\right|_{R_{\rho}}}
\sum_{\left.(m_{21}\geq m_{22})\right|_{R_{\rho}}}\cdots\sum_{\left.(m_{N1}\geq m_{N2}\cdots\geq m_{N-1N-1})\right|_{R_{\rho}}}t^{
\Delta(m)}\prod_{i=1}^{N-1} z_i^{\sum_{j} m_{ij}}\left.\left(\prod_{k=1}^{N-1} P_{U(k)}(m,t)\right)\right|_{R_{\rho}}, \label{formula}
\end{align}
where the summations are modified in a way prescribed by a restriction map $R_{\rho}$ associated to the frozen color D3-branes. $z_i, i=1, \cdots, N-1$ are fugacities for the non-perturbative $SU(N)$ topological symmetry associated to the Coulomb branch moduli. We will now define this map, and explain how it is determined by the partition $\rho$.

Let us label the cells between adjacent NS5-branes of the brane diagram as $1$, $2$, $\cdots$, $N$, starting from the leftmost cell. 
From the brane picture, the restriction map $R_{\rho}$ associated to a partition of the type $\rho=[a_1,\cdots a_n]$ with $a_1 \geq a_2 \geq \cdots \geq a_n$ and $\sum_{i=1}^n a_i = N$ can be read off as follows:
\begin{itemize}
\item The restriction on the magnetic charges.

From the quiver theory in \eqref{generalquiver}, the total number of color D3-branes which are frozen in the $k$-th cell is given by \eqref{Nk}, namely
\begin{equation}
N_k=\sum_{i=1}^n\left(a_i-(N-k)\right)H\left(a_i-(N-k)\right). \label{numberfrozen}
\end{equation}
Hence, $N_k$ magnetic charges among the $k$ magnetic charges of $U(k)$ are set to zero in the summations of \eqref{formula}. 

$N_k$ is always smaller than $k$ except for the case where there is no Coulomb branch moduli. Then we have several ways to choose $N_k$ magnetic charges which we set to zero among the $k$ magnetic charges in the $k$-th cell. The rule is that we consider all the possible choices which are compatible with the condition for the magnetic charges to remain in the same Weyl chamber $\Gamma(^LG)/\mathcal{W}_{^LG}$.

\item The change of the factor $P_{U(k)}$

The factor $P_{U(k)}$ should be composed of non-frozen Coulomb branch moduli. Therefore, in the $k$-th cell, the factor $P_{U(k)}$ is replaced with $P_{U\left(k-N_k\right)}$ with the $N_k$ defined in \eqref{numberfrozen}. 
\end{itemize}

In this way, we propose that the restriction rule gives the Hilbert series for the Coulomb branch part $\mathcal{C}_{\rho}$ of the mixed branch $\mathcal{M}_{\rho}$. Furthermore, by using the product structure of the mixed branch of the $T[SU(N)]$ theory \eqref{mixedTSUN1}, 
\be
\mathcal{M}_{\rho} = \mathcal{C}_{\rho} \times \mathcal{C}_{\rho^D},
\ee
the Hilbert series for the mixed branch $\mathcal{M}_{\rho}$ can be written by
\be
HS_{\mathcal{M}_{\rho}}\left(t, z_i, y_j\right) = HS_{\rho}\left(t,z_i\right) \times HS_{\rho^D}\left(t, \prod_{j=1}^{N-1}x_j^{M_{ij}}\right), \label{HSformixed}
\ee
where $x_j, j=1, \cdots, N-1$ are the fugacities associated to the perturbative $SU(N)$ flavor symmetry. $M_{ij}$ is an element of a matrix yielding a linear combination of the Cartan generators of the flavor symmetry group, depending on the definition of the fugacities. 

In the Hilbert series computation for the Coulomb branch, we will use the fugacities $z_i, i=1, \cdots, N-1$ associated to the Cartan generators $H_i^z$ which give charges for the simple roots of the $SU(N)$\footnote{The simple roots of the $su(N)$ Lie algebra can be expressed as $e_i - e_{i+1}, i=1, \cdots, N-1$ where $e_i, i=1, \cdots, N$ are orthonormal bases in $\mathbb{R}^N$.} as
\be
H^z_i |e_j - e_{j+1}\rangle = \delta_{ij}|e_j - e_{j+1}\rangle, \label{basisz}
\ee
for $i, j=1, \cdots, N-1$. On the other hand, for the Hilbert series computation for the Higgs branch, we will use the fugacities $x_i, i=1, \cdots, N-1$ associated to the Cartan generators $H_i^x$ which give charges to the simple roots of the $su(N)$ Lie algebra as 
\be
H^x_i |e_j - e_{j+1}\rangle = C_{ij}^{su(N)}|e_j - e_{j+1}\rangle, \label{basisx}
\ee
for $i, j=1, \cdots. N-1$ where $C_{ij}^{su(N)}$ is an element of the Cartan matrix of the $su(N)$ Lie algebra. Due to these choices of the Cartan generators,, the matrix $M_{ij}$ is in fact the Cartan matrix $C_{ij}^{su(N)}$ in the later computation which we will perform. 

Although we focus on mixed branches of the $T[SU(N)]$ theory, the restriction rule will be applicable to the computation of mixed branches of more general 3d $\mathcal{N}=4$ gauge theories which have the type IIB brane construction without orientifolds.

The similar restriction has been made use of for computing the Hilbert series of 3d $\mathcal{N}=2$ gauge theories \cite{Hanany:2015via, Cremonesi:2015dja, Cremonesi:2016nbo}. In that case, the restriction of the magnetic charges or the corresponding Coulomb branch moduli occurs due to the generation of non-perturbative superpotentials which lift a part of the Coulomb branch moduli. In the current case, the restriction of the Coulomb branch arises since we consider a sublocus of the full Coulomb branch of the $T[SU(N)]$ theory where a Higgs branch opens up. Furthermore, the restriction of the magnetic charges can be understood from the frozen D3-branes in the brane picture.

\subsection{The restriction rule with an example}

The algorithmic rule defined above is quite straightforward to apply, however it can seem involved at first. Hence let us give now an explicit example of how the rule should be applied to determine the frozen magnetic charges, in a nontrivial case of the partition $[3,2]$. In this case $N=5$ and $n=2$. Then,
\begin{itemize}
\item For $a_1=3$, the restriction appears from the $3$rd cell since $a_1 - (5-k) > 0$ when $k \geq 3$. Then, 
\begin{enumerate}
\item For $k=3$, in the $3$rd cell we set to zero $a_1 - (5-k) = 1$ magnetic charge.
\item For $k=4$, in the $4$th cell we set to zero $a_1-(5-k)=2$ magnetic charges.
\end{enumerate}
\item For $a_2=2$, the restriction appears from the $4$th cell since $a_2-(5-k) > 0$ when $k \geq 4$. Then, 
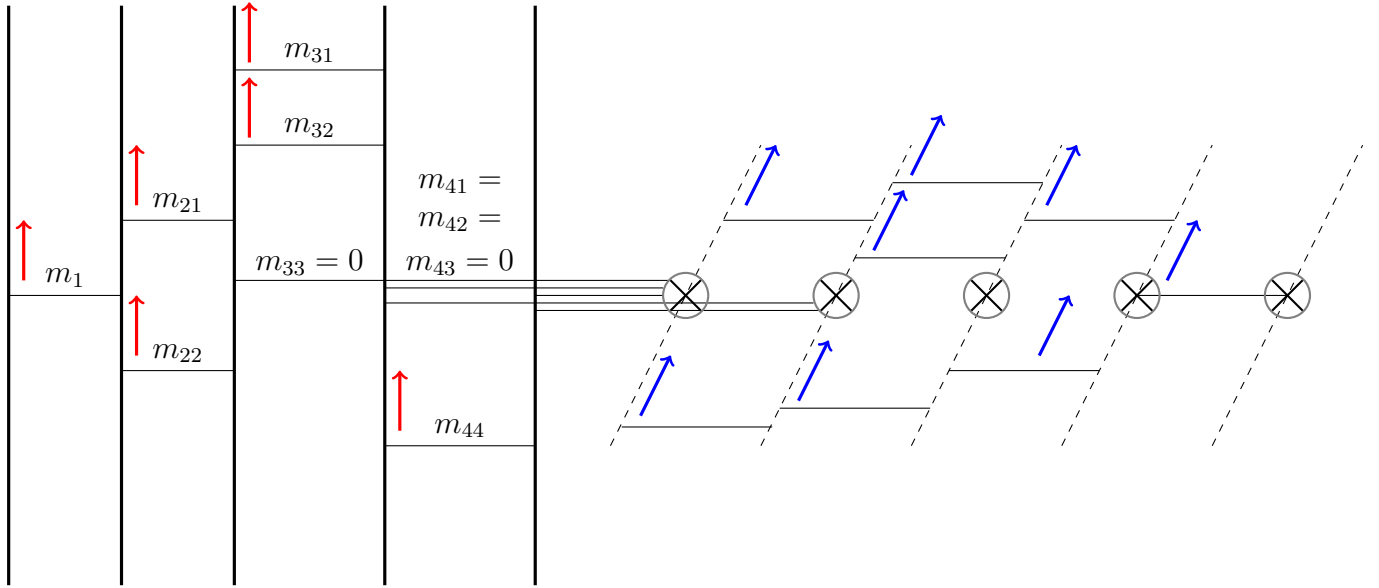
\begin{enumerate}
\item For $k=4$, in the $4$th cell we set to zero $a_2-(5-k)=1$ magnetic charge. 
\end{enumerate}
\end{itemize}
Therefore, in this case, we see that a total of $2+1=3$ magnetic charges must be put to zero in the $4$th cell, and only $1$ magnetic charge should be put to zero in the $3$rd cell. This information can be also understood in a clear way from the brane picture of the $[3,2]$ branch, as shown in the figure \ref{frozenexplain}, where one color D3-brane is fixed in the $3$rd cell and three color D3-branes are frozen in the $4$th cell.

\begin{figure}[t]
\begin{center}
\begin{tikzpicture}
\node (u3u) at ( -6,4)  {};
\node (u3d) at ( -6,-4)  {}
edge [lpre] node[auto,swap] {} (u3u);
\node (u5u) at ( -4.5,4)  {};
\node (u5d) at ( -4.5,-4)  {}
edge [lpre] node[auto,swap] {} (u5u);
\node (u4u) at ( -3,4)  {};
\node (u4d) at ( -3,-4)  {}
edge [lpre] node[auto,swap] {} (u4u);
\node (u1u) at ( -1,4)  {};
\node (u1d) at ( -1,-4)  {}
edge [lpre] node[auto,swap] {} (u1u);
\node (u2u) at ( 1,4)  {};
\node (u2d) at ( 1,-4)  {}
edge [lpre] node[auto,swap] {} (u2u);

\draw (-6,0)--(-4.5,0);

\draw (-3,0.2)--(2.8,0.2);
\draw (-1,0.1)--(2.7,0.1);
\draw (1,0)--(2.7,0);
\draw (-1,-0.1)--(4.7,-0.1);
\draw (1,-0.2)--(4.8,-0.2);

\draw(-4.5,1)--(-3,1);
\draw(-4.5,-1)--(-3,-1);

\draw(-3,3)--(-1,3);
\draw(-3,2)--(-1,2);

\draw(-1,-2)--(1,-2);

\node [below] at (-5.25,0.5) {$m_1$};
\node [below] at (-3.75,1.5) {$m_{21}$};
\node [below] at (-3.75,-0.5) {$m_{22}$};
\node [below] at (-2,3.5) {$m_{31}$};
\node [below] at (-2,2.5) {$m_{32}$};
\node [below] at (-2,0.75) {$m_{33}=0$};
\node [below] at (0,1.75) {$m_{41}=$};
\node [below] at (0,1.25) {$m_{42}=$};
\node [below] at (0,0.75) {$m_{43}=0$};
\node [below] at (0,-1.5) {$m_{44}$};

\draw [dashed] (2,-2)--(4,2);
\draw [dashed] (4,-2)--(6,2);
\draw [dashed] (6,-2)--(8,2);
\draw [dashed] (8,-2)--(10,2);
\draw [dashed] (10,-2)--(12,2);

\draw (3.5,1)--(5.5,1);
\draw (2.15,-1.75)--(4.15,-1.75);

\draw (5.75,1.5)--(7.75,1.5);
\draw (4.25,-1.5)--(6.25,-1.5);
\draw (5.25,0.5)--(7.25,0.5);

\draw (7.5,1)--(9.5,1);
\draw (6.5,-1)--(8.5,-1);
\draw (9,0)--(11,0);

\node [DfiveBig] (d53) at (3,0) {};
\node [DfiveBig] (d53) at (5,0) {};
\node [DfiveBig] (d51) at (7,0) {};
\node [DfiveBig] (d51) at (9,0) {};
\node [DfiveBig] (d52) at (11,0) {};

\draw [->, very thick, red] (-5.8,0.2)--(-5.8,1);
\draw [->, very thick, red] (-4.3,1.2)--(-4.3,2);
\draw [->, very thick, red] (-4.3,-0.8)--(-4.3,0);
\draw [->, very thick, red] (-2.8,2.1)--(-2.8,2.9);
\draw [->, very thick, red] (-2.8,3.1)--(-2.8,3.9);
\draw [->, very thick, red] (-0.8,-1.8)--(-0.8,-1);

\draw [->, very thick, blue] (3.8,1.2)--(4.2,2);
\draw [->, very thick, blue] (2.4,-1.6)--(2.8,-0.8);
\draw [->, very thick, blue] (6,1.6)--(6.4,2.4);
\draw [->, very thick, blue] (5.5,0.6)--(5.9,1.4);
\draw [->, very thick, blue] (4.5,-1.4)--(4.9,-0.6);
\draw [->, very thick, blue] (7.7,-0.8)--(8.1,0);
\draw [->, very thick, blue] (7.8,1.2)--(8.2,2);
\draw [->, very thick, blue] (9.4,0.2)--(9.8,1);

\node [below] at (4.4,-5) {$= k.\qquad\;\;\;$ Number of the cell};
\node [below] at (5,-6) {$= N_k.\qquad\;$ Number of frozen branes};
\node [below] at (5.1,-7) {$= k-N_k.\;\;$ Number of mobile branes.};

\node [below] at (-5.25,-5) {$1$};
\node [below] at (-3.75,-5) {$2$};
\node [below] at (-2,-5) {$3$};
\node [below] at (0,-5) {$4$};

\node [below] at (-5.25,-6) {$0$};
\node [below] at (-3.75,-6) {$0$};
\node [below] at (-2,-6) {$1$};
\node [below] at (0,-6) {$3$};

\node [below] at (-5.25,-7) {$1$};
\node [below] at (-3.75,-7) {$2$};
\node [below] at (-2,-7) {$2$};
\node [below] at (0,-7) {$1$};

\end{tikzpicture}
\caption{The $[3,2]$ example, and the different numbers of frozen branes in every cell.}
\label{frozenexplain}
\end{center}
\end{figure}

Now, in the $4$th cell we have $4$ magnetic charges in total. Let's call the $m_{41},m_{42},m_{43},m_{44}$ and they are subject to be in the the same Weyl chamber of the weight space of $U(4)$, therefore they satisfy
\begin{equation}
m_{41}\geq m_{42}\geq m_{43}\geq m_{44}. \label{weyl1}
\end{equation}
Among them we should choose three to vanish and the rule is that we must take into account all the possible ways. By looking at the Weyl chamber condition \eqref{weyl1}, we see that there are only two ways. We can have
\begin{enumerate}
\item $0=m_{41}= m_{42}= m_{43}\geq m_{44}$,
\item $m_{41}\geq m_{42}= m_{43}= m_{44}=0$.
\end{enumerate}

A similar reasoning works also for the magnetic charge that should be set to zero in the $3$rd cell.
In the $3$rd cell there are three magnetic charges $m_{31}, m_{32}, m_{33}$ for $U(3)$ satisfying
\begin{equation}
m_{31}\geq m_{32} \geq m_{33},
\end{equation}
and we see that in this case we have three ways to put one of the magnetic charges to zero,
\begin{enumerate}
\item $0=m_{31}\geq m_{32}\geq m_{33}$,
\item $m_{31}\geq m_{32}= 0\geq m_{33}$,
\item $m_{31}\geq m_{32}\geq m_{33}=0$.
\end{enumerate}

Therefore, in this example, we see there are in total $3\times 2$ different sets of magnetic charges that need to be put to zero, and therefore the Hilbert series of the full Coulomb branch will split in six different sub-sums, depending on the way in which the non-zero charges are chosen. In the restriction of the Hilbert series, one has to take into account all of these conditions and sum over all of them. However, to avoid oversumming, if some value for the magnetic charge is repeated, it should be counted only once. For example, we see that $m_{31}=m_{32}=m_{33}=m_{34}=0$ is repeated both in the first and the second way for the $4$th cell.

For the practical computation of the restriction of the magnetic charges, we can divide the possibilities of setting which magnetic charges to zero into disjoint sets. This will crucially avoid the overcounting problem outlined above. Let us consider a gauge node $U(k)$ with the magnetic charges satisfying the Weyl chamber condition
\begin{equation}
m_1 \geq m_2 \geq \cdots \geq m_k.
\end{equation}
In a Coulomb branch part of a mixed branch, the rule says that $N_k$ of the magnetic charges are zero. Then, there are $k-N_k+1$ possibilities of which $N_k$ magnetic charges are zero. For $i=0, \cdots, k-N_k$, we can consider the following set
\begin{equation}
m_1 \geq \cdots \geq m_{i-1} \geq m_{i} > 0 \geq m_{i+N_k+1} \geq m_{i+N_k+2} \geq \cdots \geq m_k. \label{disjointsets}
\end{equation}
with 
\be
m_{i+1} = m_{i+2} = \cdots = m_{i+N_k} = 0.
\ee
These sets are all disjoint between each other for all $i=0, \cdots, k-N_k$ and in fact the sum of the sets exhausts all the elements in the summation after the restriction. Hence, in the practical calculation one can use the disjoint sets \eqref{disjointsets} to sum up all the possibilities of the restriction of the magnetic charges.

\section{Coulomb Branch Examples}
\label{section5}

In this section we will work out explicitly some examples of the general procedure outlined in section \ref{sec:restriction}, in order to explain the rather abstract rule that defines the restriction map in terms of the partition $\rho$ and perform some explicit checks that our conjecture holds. 

\subsection{The case of $[2, 1]$ of $T[SU(3)]$} 
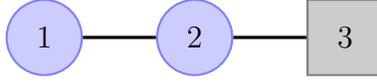
\begin{figure}[t]
\begin{center}
\begin{tikzpicture}
\node (m1) at ( -2,0) [gauge] {$1$};
\node (m2) at ( 0,0) [gauge] {$2$}
edge [lpre] node[auto,swap] {} (m1);
\node (n) at ( 2,0) [global] {$3$}
edge [lpre] node[auto,swap] {} (m2);
\end{tikzpicture}
\caption{The quiver graph for $T[SU(3)]$ theory.\label{easy}}
\end{center}
\end{figure}

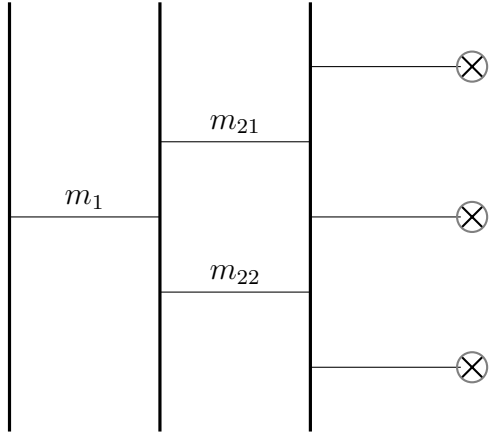
\begin{figure}[t]
\begin{center}
\begin{tikzpicture}
\node (u1u) at ( 0,3)  {};
\node (u1d) at ( 0,-3)  {}
edge [lpre] node[auto,swap] {} (u1u);
\node (u2u) at ( 2,3)  {};
\node (u2d) at ( 2,-3)  {}
edge [lpre] node[auto,swap] {} (u2u);
\node (u3u) at ( 4,3)  {};
\node (u3d) at ( 4,-3)  {}
edge [lpre] node[auto,swap] {} (u3u);
\draw (0,0)--(2,0);
\draw (2,1)--(4,1);
\draw (2,-1)--(4,-1);
\draw (4,2)--(6,2);
\draw (4,0)--(6,0);
\draw (4,-2)--(6,-2);
\node [below] at (1,0.5) {$m_1$};
\node [below] at (3,1.5) {$m_{21}$};
\node [below] at (3,-0.5) {$m_{22}$};
\node [Dfive] (d51) at (6.15,2) {};
\node [Dfive] (d52) at (6.15,0) {};
\node [Dfive] (d53) at (6.15,-2) {};
\end{tikzpicture}
\caption{The brane picture for the $T[SU(3)]$ theory yielding $U(1)-U(2)-[3]$.}
\label{brane123}
\end{center}
\end{figure}

To begin, let us think of the easiest possible case. We consider the $T[SU(3)]$ theory defined by the following linear quiver of figure \ref{easy}.
The brane picture is given in figure \ref{brane123} where $m_1$ is the magnetic charge for $U(1)$ and $m_{21}, m_{22}$ are the magnetic charges for $U(2)$ which satisfy $m_{21} \geq m_{22}$. 
The Hilbert series for the full Coulomb branch of is given by the general formula \eqref{HSC}
\begin{align}
HS(t,z_1,z_2)=\sum_{m_1=-\infty}^{\infty}\sum_{m_{21}\geq m_{22}} t^{\Delta(m_{1},m_{21},m_{22})}z_1^{m_1}z_2^{m_{21}+m_{22}}P_{U(1)}(m_1,t)P_{U(2)}(m_{21},m_{22},t),
\label{fullCou21}
\end{align}
where the dimension formula \eqref{DimensionFormula} reads 
\begin{equation}
\Delta(m_1,m_{21},m_{22})=-|m_{21}-m_{22}|+\dfrac{1}{2}(|m_{21}-m_1|+|m_{22}-m_1|+3|m_{21}|+3|m_{22}|),
\end{equation}
and the classical factors are 
\be
P_{U(1)}(m_1, t) =\frac{1}{1-t},
\ee
and
\be
P_{U(2)}(m_{21}, m_{22}, t) = \left\{
\begin{aligned}&\frac{1}{(1-t)(1-t^2)},\quad \text{for}\;m_{21}=m_{22},\\
&\frac{1}{(1-t)^2},\qquad\qquad \text{for}\;m_{21} > m_{22}.
\end{aligned}
\right.
\ee

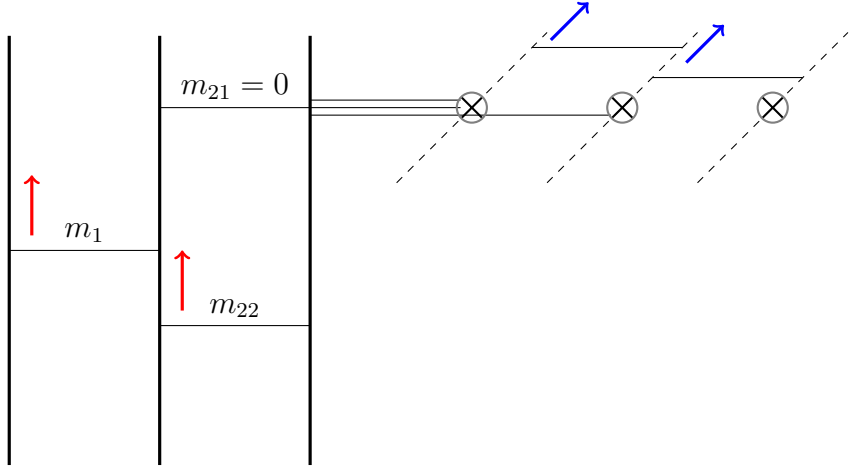
\begin{figure}[t]
\begin{center}
\begin{tikzpicture}
\node (u1u) at ( 0,3)  {};
\node (u1d) at ( 0,-3)  {}
edge [lpre] node[auto,swap] {} (u1u);
\node (u2u) at ( 2,3)  {};
\node (u2d) at ( 2,-3)  {}
edge [lpre] node[auto,swap] {} (u2u);
\node (u3u) at ( 4,3)  {};
\node (u3d) at ( 4,-3)  {}
edge [lpre] node[auto,swap] {} (u3u);
\draw (0,0)--(2,0);
\draw (2,1.9)--(4,1.9);
\draw (2,-1)--(4,-1);
\draw (4,2)--(6,2);
\draw (4,1.8)--(8,1.8);
\draw (8.55,2.3)--(10.55,2.3);
\draw (4,1.9)--(6,1.9);
\draw [dashed] (5.15,0.9)--(7.15,2.9);
\draw [dashed] (7.15,0.9)--(9.15,2.9);
\draw [dashed] (9.15,0.9)--(11.15,2.9);
\draw (6.95,2.7)--(8.95,2.7);
\node [below] at (1,0.5) {$m_1$};
\node [below] at (3,2.5) {$m_{21}=0$};
\node [below] at (3,-0.5) {$m_{22}$};
\node [Dfive] (d51) at (6.15,1.9) {};
\node [Dfive] (d52) at (8.15,1.9) {};
\node [Dfive] (d53) at (10.15,1.9) {};
\draw [->, very thick, blue] (7.2,2.8)--(7.7,3.3);
\draw [->, very thick, red] (0.3,0.2)--(0.3,1);
\draw [->, very thick, red] (2.3,-0.8)--(2.3,0);
\draw [->, very thick, blue] (9,2.5)--(9.5,3);
\end{tikzpicture}
\caption{The brane picture realizing the mixed branch $\rho=[2,1]$. In this first case the restriction amounts to take $m_{21}=0 \geq m_{22}$.}
\label{mzero}
\end{center}
\end{figure}

Then we focus on the mixed branch $\rho=[2,1]$. In order to satisfy the s-rule \cite{Hanany:1996ie} we must set the position of one of the two D3-branes in the second cell to be exactly equal to one of the mass parameters, and therefore equal to the position of one of the flavor D5-brane. Computationally, this is implemented by setting to zero the magnetic charge associated to the position of that brane. Figure \ref{mzero} shows how the brane system looks for the mixed branch of $\rho=[2,1]$. 

\begin{figure}[t]
\begin{center}
\begin{tikzpicture}
\node (u1u) at ( 0,3)  {};
\node (u1d) at ( 0,-3)  {}
edge [lpre] node[auto,swap] {} (u1u);
\node (u2u) at ( 2,3)  {};
\node (u2d) at ( 2,-3)  {}
edge [lpre] node[auto,swap] {} (u2u);
\node (u3u) at ( 4,3)  {};
\node (u3d) at ( 4,-3)  {}
edge [lpre] node[auto,swap] {} (u3u);
\draw (0,0)--(2,0);
\draw (2,2)--(4,2);
\draw (2,-1)--(4,-1);
\draw (4,-0.9)--(7.9,-0.9);
\draw (4,-1)--(6,-1);
\draw (4,-1.1)--(6,-1.1);
\draw [dashed] (5.15,-2)--(7.15,0);
\draw [dashed] (7.05,-2)--(9.05,0);
\draw [dashed] (9.05,-2)--(11.05,0);
\draw (6.85,-0.3)--(8.75,-0.3);
\draw (7.25,-1.8)--(9.25,-1.8);

\node [below] at (1,0.5) {$m_1$};
\node [below] at (3,2.5) {$m_{21}$};
\node [below] at (3,-0.5) {$m_{22}=0$};

\node [Dfive] (d51) at (6.15,-1) {};
\node [Dfive] (d52) at (8.05,-1) {};
\node [Dfive] (d53) at (10.05,-1) {};

\draw [->, very thick, red] (0.3,0.1)--(0.3,0.9);
\draw [->, very thick, red] (2.3,2.1)--(2.3,2.9);
\draw [->, very thick, blue] (7.3,-0.2)--(7.8,0.3);
\draw [->, very thick, blue] (8,-1.7)--(8.5,-1.2);
\end{tikzpicture}
\caption{Another brane picture for $\rho=[2,1]$. In this second case the restriction amounts to take $m_{21} > m_{22}=0$.}
\label{mzero2}
\end{center}
\end{figure}
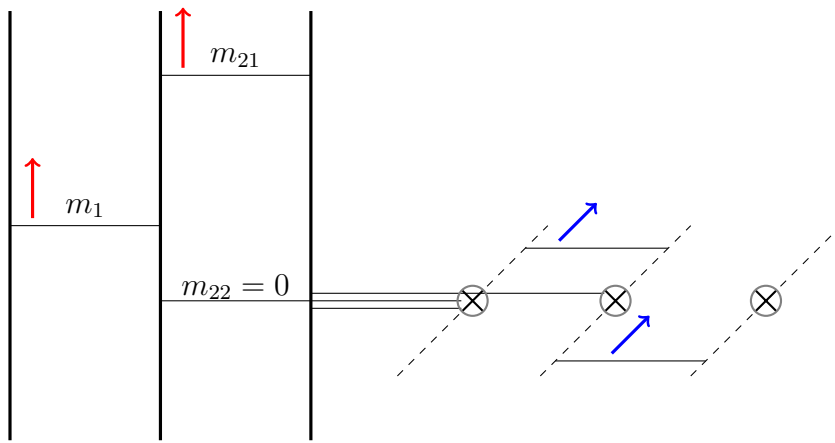

Now we should point out that there are two ways to set one of the two gauge branes to zero: one is  $m_{21}=0\geq m_{22}$, as shown in figure \ref{mzero} and the other is to set $m_{21}> m_{22}=0$, as shown in figure \ref{mzero2}.
Both these cases are allowed and we should sum over both of them. With this we mean that the Hilbert series for the Coulomb branch part of mixed branch $\rho=[2,1]$ will be given by the Hilbert series of the full Coulomb branch \eqref{fullCou21} in which $m_{21}=0 \geq m_{22}$ plus the Hilbert series of the full Coulomb branch \eqref{fullCou21} in which $m_{21} > m_{22}=0$. In this addition, we only count the magnetic charge $m_{21}=m_{22}=0$ once and there is no overcounting. 

By the second rule in section \ref{sec:rule}, the map $R_{[2,1]}$
also restricts the classical factors, replacing $P_{U(2)}$ with $P_{U(1)}$. The physical intuition for this fact is that since one of the two branes is frozen to a specific position, the residual gauge group becomes $U(1)$. Therefore, the classical dressing factor will be reduced to $P_{U(1)}$ from $P_{U(2)}$.

The Hilbert series for the Coulomb branch part of the $[2,1]$ mixed branch is therefore given by
\begin{equation}
\begin{aligned}
HS(t,z_1,z_2)&=\sum_{m_1=-\infty}^{\infty}\sum_{m_{21}=0, m_{22}\leq 0}t^{\Delta_1(m_1,m_{22})}z_1^{m_1}z_2^{m_2}P_{U(1)}(m_1,t)P_{U(1)}(m_{22},t)\\
&+\sum_{m_1=-\infty}^{\infty}\sum_{m_{22}=0, m_{21}> 0}t^{\Delta_2(m_1,m_{21})}z_1^{m_1}z_2^{m_2}P_{U(1)}(m_1,t)P_{U(1)}(m_{21},t),
\end{aligned}
\label{21restr}
\end{equation}
where $\Delta_1(m_1,m_{22})=\Delta(m_1,0,m_{22})$ and $\Delta_2(m_1,m_{21})=\Delta(m_1,m_{21},0)$.

In detail, by performing this truncation of the sum at the ninth order in $t$, we get\footnote{This computation, writing the HS in terms of characters, was performed explicitly only up to order $t^{9}$. However, the obtained result strongly implies that the same structure will continue to higher orders. Therefore we conjecture that the Hilbert series is given by
\be
HS(z,t)=\sum_{k=0}^{\infty}[k,k]_zt^k.
\ee.}
\begin{equation}
HS(z,t)=\sum_{k=0}^{9}[k,k]_zt^k + \mathcal{O}(t^{10}),
\label{21restr}
\end{equation}
where $[n_1, n_2]_z$ is the character of the representation $[n_1, n_2]$ where $n_1, n_2$ are Dynkin labels of the $su(3)$ Lie algebra\footnote{Here and everywhere else in section \ref{section5} and \ref{sec:Higgs} we use a Dynkin label notation for the characters of a representation of a Lie algebra. For example, $[2]$ means the character of the adjoint of $\mathfrak{su}(2)$. By the basis of the Cartan generators in \eqref{basisz}, the character is given by $[2]_z=z +1+z^{-1}$. On the other hand, by the basis of the Cartan generators in \eqref{basisx}, the character is given by $[2]_x=x^2 +1+x^{-2}$.}. Since the Hilbert series of \eqref{21restr} is written by the characters of the $su(3)$ Lie algebra, it implies that the topological symmetry is enhanced to $SU(3)$. 

We now want to check that the restriction rule indeed works. We compare the Hilbert series that we obtained by restricting the summation over the magnetic charges, with the Hilbert series of the full Coulomb branch of the $T^{[2,1]}[SU(N)]$ theory. To do so we first go to the IR theory, effectively giving infinite vev to the scalars parameterizing the Higgs branch. The resulting brane configuration after a sequence of Hanany-Witten transitions was already obtained in figure \ref{IR1111}, yielding the $[1]-U(1)-U(1)-[1]$ linear quiver theory. For this theory the monopole dimension is
\begin{align}
\Delta(n_1, n_2)=\dfrac{1}{2}\left(\left|n_1\right|+\left|n_2-n_1\right|+\left|n_2\right|\right),
\end{align}
where $n_1, n_2$ are the magnetic charges of the two $U(1)$'s. The Hilbert series of the full Coulomb branch is given by
\begin{align}
HS(t,z_1,z_2)=\sum_{n_1=-\infty}^{\infty}\sum_{n_2=-\infty}^{\infty}t^{\Delta(n_1,n_2)}z_1^{n_1}z_{2}^{n_2}P_{U(1)}(n_1,t)P_{U(1)}(n_2,t).
\end{align}
Performing this computation explicitly gives us
\begin{equation}
HS(z,t)=\sum_{k=0}^{9}[k,k]_zt^k + \mathcal{O}(t^{10}),
\label{21check}
\end{equation}
and we see that this exactly matches with the equation \eqref{21restr}. This matching was checked at order $30$ in $t$.

\subsection{Other explicit checks}
\label{sec:Coulombexs}
We then exemplify the restriction rules in section \ref{sec:restriction} by more non-trivial examples. 

\subsubsection{The mixed branch $\rho=[2,2]$}
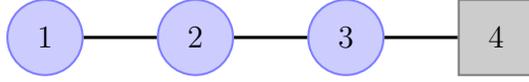
\begin{figure}[t]
\begin{center}
\begin{tikzpicture}
\node (m1) at ( -3,0) [gauge] {$1$};
\node (m2) at ( -1,0) [gauge] {$2$}
edge [lpre] node[auto,swap] {} (m1);
\node (m3) at ( 1,0) [gauge] {$3$}
edge [lpre] node[auto,swap] {} (m2);
\node (f1) at ( 3,0) [global] {$4$}
edge [lpre] node[auto,swap] {} (m3);
\end{tikzpicture}
\caption{The quiver graph for the $T[SU(4)]$ theory.\label{QuiverSU4}}
\end{center}
\end{figure}

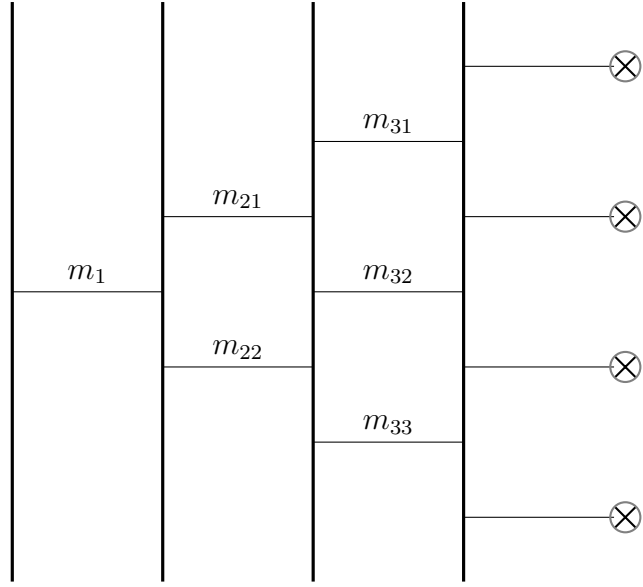
\begin{figure}[t]
\begin{center}
\begin{tikzpicture}
\node (u1u) at ( 0,4)  {};
\node (u1d) at ( 0,-4)  {}
edge [lpre] node[auto,swap] {} (u1u);
\node (u2u) at ( 2,4)  {};
\node (u2d) at ( 2,-4)  {}
edge [lpre] node[auto,swap] {} (u2u);
\node (u3u) at ( 4,4)  {};
\node (u3d) at ( 4,-4)  {}
edge [lpre] node[auto,swap] {} (u3u);
\node (u4u) at ( 6,4)  {};
\node (u4d) at ( 6,-4)  {}
edge [lpre] node[auto,swap] {} (u4u);
\draw (0,0)--(2,0);
\draw (2,1)--(4,1);
\draw (2,-1)--(4,-1);
\draw (4,2)--(6,2);
\draw (4,0)--(6,0);
\draw (4,-2)--(6,-2);
\draw (6,1)--(8,1);
\draw (6,-1)--(8,-1);
\draw (6,-3)--(8,-3);
\draw (6,3)--(8,3);
\node [below] at (1,0.5) {$m_1$};
\node [below] at (3,1.5) {$m_{21}$};
\node [below] at (3,-0.5) {$m_{22}$};
\node [below] at (5,2.5) {$m_{31}$};
\node [below] at (5,0.5) {$m_{32}$};
\node [below] at (5,-1.5) {$m_{33}$};
\node [Dfive] (d51) at (8.15,1) {};
\node [Dfive] (d52) at (8.15,-1) {};
\node [Dfive] (d53) at (8.15,-3) {};
\node [Dfive] (d53) at (8.15,3) {};
\end{tikzpicture}
\caption{The brane picture for the $T[SU(4)]$ theory yielding the linear quiver $U(1)-U(2)-U(3)-[4]$.}
\label{brane1234}
\end{center}
\end{figure}

In this example we start by considering now the $3d$ $\mathcal{N}=4$ $T[SU(4)]$ theory given by the quiver diagram depicted in figure \ref{QuiverSU4}.
This theory can be also realized in terms of the brane picture in figure \ref{brane1234} where $m_1$ is the magnetic charge of the $U(1)$, $m_{21}, m_{22}$ are the magnetic charges of the $U(2)$, $m_{31}, m_{32}, m_{33}$ are the magnetic charges of the $U(3)$ and $m_{41}, m_{42}, m_{43}, m_{44}$ are the magnetic charges of the $U(4)$.
The monopole dimension formula for the full Coulomb branch reads:
\begin{equation}
\begin{aligned}
\Delta\left(\vec{m}\right)&=-|m_{21}-m_{22}|-|m_{31}-m_{32}|-|m_{31}-m_{33}|-|m_{32}-m_{33}|\\
&+\dfrac{1}{2}\left(|m_{21}-m_{11}+|m_{22}-m_{11}|+|m_{31}-m_{21}|+|m_{31}-m_{22}|+|m_{32}-m_{21}|\right.\\
&\left.+|m_{32}-m_{22}|+|m_{33}-m_{21}|+|m_{33}-m_{22}|+4|m_{31}|+4|m_{32}|+4|m_{33}|\right).
\end{aligned}
\end{equation}
with the magnetic charges $\vec{m}=\left(m_1. m_{21}, m_{22}, m_{31}, m_{32}, m_{33}\right)$ satisfying $m_{21} \geq m_{22}$ and $m_{31} \geq m_{32} \geq m_{33}$.

The Hilbert series for the full Coulomb branch of this theory is
\begin{equation}
\begin{aligned}
HS(z_,t)&:=\sum_{m_1=-\infty}^{\infty}\sum_{m_{21}\geq m_{22}}\sum_{m_{31}\geq m_{32}\geq m_{33}}t^{\Delta(m_1,m_{21},m_{22},m_{31},m_{32},m_{33})}\\
&\cdot P_{U(1)}(m_1,t) P_{U(2)}(m_{21},m_{22},t) P_{U(3)}(m_{31},m_{32},m_{33},t)z_1^{m_1}z_2^{\left(m_{21}+m_{22}\right)}z_3^{\left(m_{31}+m_{32}+m_{33}\right)}. \label{fullCoulombTSU4}
\end{aligned}
\end{equation}

Now we are interested in studying the mixed branch given by the partition $\rho=[2,2]$. The first rule in section \ref{sec:rule} says that we can set two magnetic charges to zero in the 3rd cell. Furthermore, one sees again that there are two different ways to set to zero two magnetic charges in the third cell: one can choose $m_{31}=m_{32}=0$, as in figure \ref{brane1234A}, or one can choose $m_{32}=m_{33}=0$ as in figure \ref{brane1234B}. 
\begin{figure}[t]
\begin{center}\begin{tikzpicture}
\node (u1u) at ( 0,4)  {};
\node (u1d) at ( 0,-4)  {}
edge [lpre] node[auto,swap] {} (u1u);
\node (u2u) at ( 2,4)  {};
\node (u2d) at ( 2,-4)  {}
edge [lpre] node[auto,swap] {} (u2u);
\node (u3u) at ( 4,4)  {};
\node (u3d) at ( 4,-4)  {}
edge [lpre] node[auto,swap] {} (u3u);
\node (u4u) at ( 6,4)  {};
\node (u4d) at ( 6,-4)  {}
edge [lpre] node[auto,swap] {} (u4u);
\draw (4,-3)--(6,-3);
\draw (0,0)--(2,0);
\draw (2,1)--(4,1);
\draw (2,-2)--(4,-2);

\draw (4,-0.9)--(9.7,-0.9);
\draw (4,-1)--(7.7,-1);
\draw (6,-1.1)--(9.7,-1.1);
\draw (6,-1.2)--(7.8,-1.2);

\draw [dashed] (7,-2)--(9,0);
\draw [dashed] (9,-2)--(11,0);
\draw [dashed] (11,-2)--(13,0);
\draw [dashed] (13,-2)--(15,0);

\draw (8.5,-0.5)--(10.5,-0.5);
\draw (10.7,-0.3)--(12.7,-0.3);
\draw (9.2,-1.8)--(11.2,-1.8);

\draw (12.25,-1)--(13.7,-1);

\node [below] at (1,0.5) {$m_1$};
\node [below] at (3,1.5) {$m_{21}$};
\node [below] at (3,-1.5) {$m_{22}$};
\node [below] at (5,0.2) {$m_{31}=$};
\node [below] at (5,-0.2) {$m_{32}=0$};
\node [below] at (5,-2.5) {$m_{33}$};
\node [DfiveBig] (d51) at (8,-1) {};
\node [DfiveBig] (d52) at (10,-1) {};
\node [DfiveBig] (d53) at (12,-1) {};
\node [DfiveBig] (d54) at (14,-1) {};

\draw [->, very thick, red] (0.2,0.2)--(0.2,1);
\draw [->, very thick, red] (2.2,1.2)--(2.2,2);
\draw [->, very thick, red] (2.2,-1.8)--(2.2,-1);
\draw [->, very thick, red] (4.2,-2.8)--(4.2,-2);

\draw [->, very thick, blue] (9,-0.3)--(9.5,0.2);
\draw [->, very thick, blue] (11.2,-0.1)--(11.7,0.4);
\draw [->, very thick, blue] (10.5,-1.7)--(11,-1.2);
\draw [->, very thick, blue] (12.6,-0.8)--(13.1,-0.3);
\end{tikzpicture}
\caption{The brane picture for the $\rho=[2,2]$ mixed branch. In this subcase, $m_{31}=m_{32}=0 \geq m_{33}$.}
\label{brane1234A}
\end{center}
\end{figure}
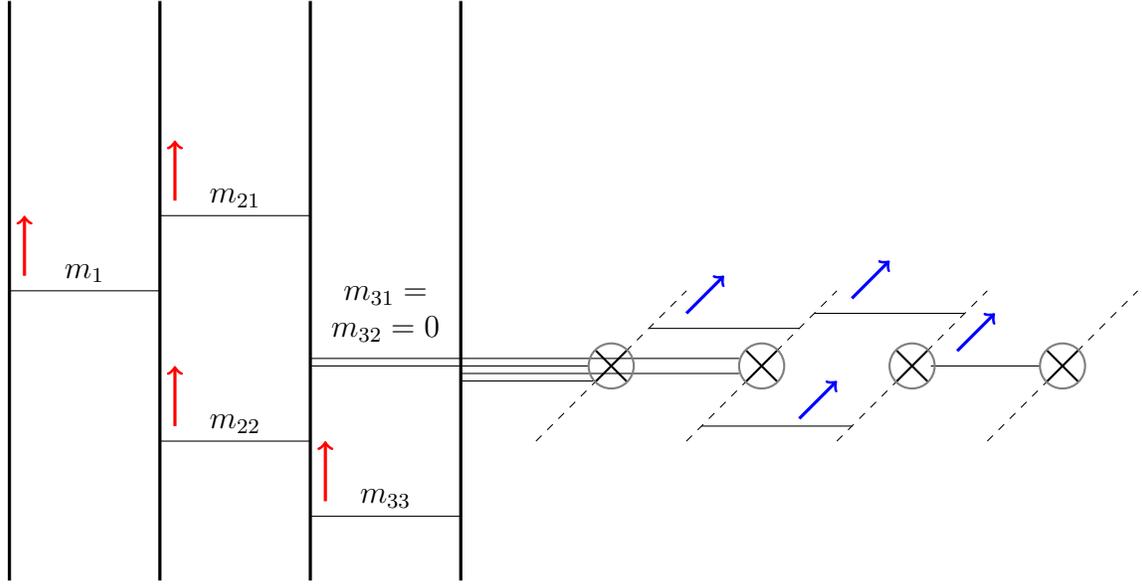
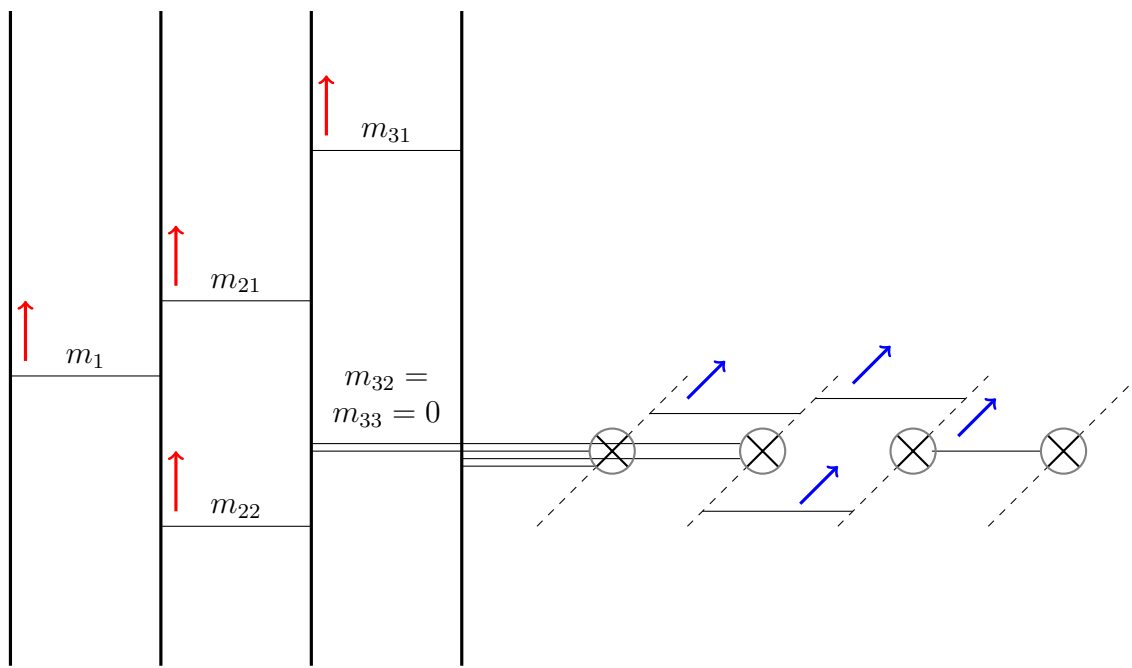
\begin{figure}[t]
\begin{center}
\begin{tikzpicture}
\node (u1u) at ( 0,5)  {};
\node (u1d) at ( 0,-4)  {}
edge [lpre] node[auto,swap] {} (u1u);
\node (u2u) at ( 2,5)  {};
\node (u2d) at ( 2,-4)  {}
edge [lpre] node[auto,swap] {} (u2u);
\node (u3u) at ( 4,5)  {};
\node (u3d) at ( 4,-4)  {}
edge [lpre] node[auto,swap] {} (u3u);
\node (u4u) at ( 6,5)  {};
\node (u4d) at ( 6,-4)  {}
edge [lpre] node[auto,swap] {} (u4u);
\draw (4,3)--(6,3);
\draw (0,0)--(2,0);
\draw (2,1)--(4,1);
\draw (2,-2)--(4,-2);

\draw (4,-0.9)--(9.7,-0.9);
\draw (4,-1)--(7.7,-1);
\draw (6,-1.1)--(9.7,-1.1);
\draw (6,-1.2)--(7.8,-1.2);

\draw [dashed] (7,-2)--(9,0);
\draw [dashed] (9,-2)--(11,0);
\draw [dashed] (11,-2)--(13,0);
\draw [dashed] (13,-2)--(15,0);

\draw (8.5,-0.5)--(10.5,-0.5);
\draw (10.7,-0.3)--(12.7,-0.3);
\draw (9.2,-1.8)--(11.2,-1.8);

\draw (12.25,-1)--(13.7,-1);

\node [below] at (1,0.5) {$m_1$};
\node [below] at (3,1.5) {$m_{21}$};
\node [below] at (3,-1.5) {$m_{22}$};
\node [below] at (5,0.2) {$m_{32}=$};
\node [below] at (5,-0.2) {$m_{33}=0$};
\node [below] at (5,3.5) {$m_{31}$};
\node [DfiveBig] (d51) at (8,-1) {};
\node [DfiveBig] (d52) at (10,-1) {};
\node [DfiveBig] (d53) at (12,-1) {};
\node [DfiveBig] (d54) at (14,-1) {};

\draw [->, very thick, red] (0.2,0.2)--(0.2,1);
\draw [->, very thick, red] (2.2,1.2)--(2.2,2);
\draw [->, very thick, red] (2.2,-1.8)--(2.2,-1);
\draw [->, very thick, red] (4.2,3.2)--(4.2,4);

\draw [->, very thick, blue] (9,-0.3)--(9.5,0.2);
\draw [->, very thick, blue] (11.2,-0.1)--(11.7,0.4);
\draw [->, very thick, blue] (10.5,-1.7)--(11,-1.2);
\draw [->, very thick, blue] (12.6,-0.8)--(13.1,-0.3);
\end{tikzpicture}
\caption{The  brane picture for the $\rho=[2,2]$ mixed branch. In this subcase, $m_{31} > m_{32}=m_{33}=0$.}
\label{brane1234B}
\end{center}
\end{figure}

One can now compute the Hilbert series for the Coulomb branch part of the $\rho=[2,2]$ mixed branch, by restricting the full summation in the way explained in section \ref{sec:rule}. By doing this one finds a series
\begin{equation}
\begin{aligned}
H(z,t)&=1+[1,0,1]_z\ t+\left([2,0,2]_z+[0,2,0]_z\right)\ t^2\\
&+\left([3,0,3]_z+[1,2,1]_z\right)\ t^3+ \left([4,0,4]_z+[2,2,2]_z+[0,4,0]_z\right)\ t^4\\
&+\left([5,0,5]_z+[3,2,3]_x+[1,4,1]_z\right)\ t^5+\left([6,0,6]_z+[4,2,4]_z+[2,4,2]_z+[0,6,0]_z\right)\ t^6\\
&+\left([7,0,7]_z+[5,2,5]_z+[3,4,3]_z+[1,6,1]_z\right)\ t^7\\
&+\left([8,0,8]_z+[6,2,6]_z+[4,4,4]_z+[2,6,2]_z+[0,8,0]_z\right)\ t^8\\
&+\left([9,0,9]_z+[7,2,7]_z+[5,4,5]_z+[3,6,3]_z+[1,8,1]_z\right)\ t^9 + \mathcal{O}(t^{10}). \label{HS22}
\end{aligned}
\end{equation}
Since the Hilbert series of \eqref{HS22} is written by the characters of the $su(4)$ Lie algebra, it implies that the topological symmetry is enhanced to $SU(4)$.

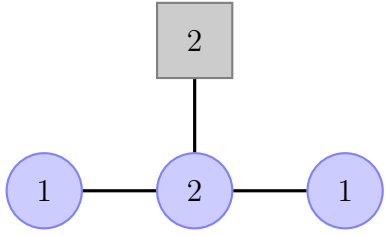
\begin{figure}[t]
\begin{center}
\begin{tikzpicture}
\node (m1) at ( -2,0) [gauge] {$1$};
\node (m2) at ( 0,0) [gauge] {$2$}
edge [lpre] node[auto,swap] {} (m1);
\node (n1) at ( 2,0) [gauge] {$1$}
edge [lpre] node[auto,swap] {} (m2);
\node (f1) at ( 0,2) [global] {$2$}
edge [lpre] node[auto,swap] {} (m2);
\end{tikzpicture}
\caption{The quiver graph for the $T^{[2,2]}[SU(4)]$ theory, obtained by sending to infinity all the unfrozen Higgs branch vevs.\label{22IR}}
\end{center}
\end{figure}

Let us then compare this result with the one obtained from the IR theory after taking the limit where the Higgs branch vevs of all the unfrozen branes become infinite. After performing some Hanany-Witten transitions, the quiver theory at the IR will be given by figure \ref{22IR}. For this theory, the dimension formula of the monopole operators is given by
\begin{equation}
\begin{aligned}
\Delta(\vec{n})&=-|n_{21}-n_{22}|+\dfrac{1}{2}\left(|n_{21}-n_1|+|n_{22}-n_1|+|n_{21}-n_2|+|n_{22}-n_2|+2|n_{21}|+2|n_{22}\right|).
\end{aligned}
\end{equation}
where $\vec{n}=\left(n_1, n_2, n_{21}, n_{22}\right)$, and $n_1, n_2$ are the magnetic charges of the two $U(1)$'s and $n_{21}, n_{22}$ are the magnetic charges of the $U(2)$.  

The Hilbert series for the IR theory is given by
\begin{equation}
\begin{aligned}
HS(z,t)&=\sum_{n_1=-\infty}^{\infty}\sum_{n_2=-\infty}^{\infty}\sum_{n_{21}\geq n_{22}}t^{\Delta(n_1,n_2,n_{21},n_{22})}z_1^{n_1}z_2^{(n_{21}+n_{22})}z_3^{n_3}P_{U(1)}\left(n_1,t\right)P_{U(1)}\left(n_2,t\right)P_{U(2)}\left(n_{21},n_{22},t\right)
\end{aligned}
\end{equation}

By computing explicitly the Hilbert series in this case we find
\begin{equation}
\begin{aligned}
H(z,t)&=1+[1,0,1]_z\ t+\left([2,0,2]_z+[0,2,0]_z\right)\ t^2\\
&+\left([3,0,3]_z+[1,2,1]_z\right)\ t^3+ \left([4,0,4]_z+[2,2,2]_z+[0,4,0]_z\right)\ t^4\\
&+\left([5,0,5]_z+[3,2,3]_x+[1,4,1]_z\right)\ t^5+\left([6,0,6]_z+[4,2,4]_z+[2,4,2]_z+[0,6,0]_z\right)\ t^6\\
&+\left([7,0,7]_z+[5,2,5]_z+[3,4,3]_z+[1,6,1]_z\right)\ t^7\\
&+\left([8,0,8]_z+[6,2,6]_z+[4,4,4]_z+[2,6,2]_z+[0,8,0]_z\right)\ t^8\\
&+\left([9,0,9]_z+[7,2,7]_z+[5,4,5]_z+[3,6,3]_z+[1,8,1]_z\right)\ t^9 + \mathcal{O}(t^{10}),
\end{aligned}
\end{equation}
which exactly agrees with \eqref{HS22}. 

\subsubsection{The mixed branch $\rho=[3,1]$}

Another example is the Coulomb branch moduli part of the mixed branch $\rho = [3, 1]$ of the 3d $T[SU(4)]$ theory. The Hilbert series for the full Coulomb branch of the $T[SU(4)]$ theory is again given by \eqref{fullCoulombTSU4}.

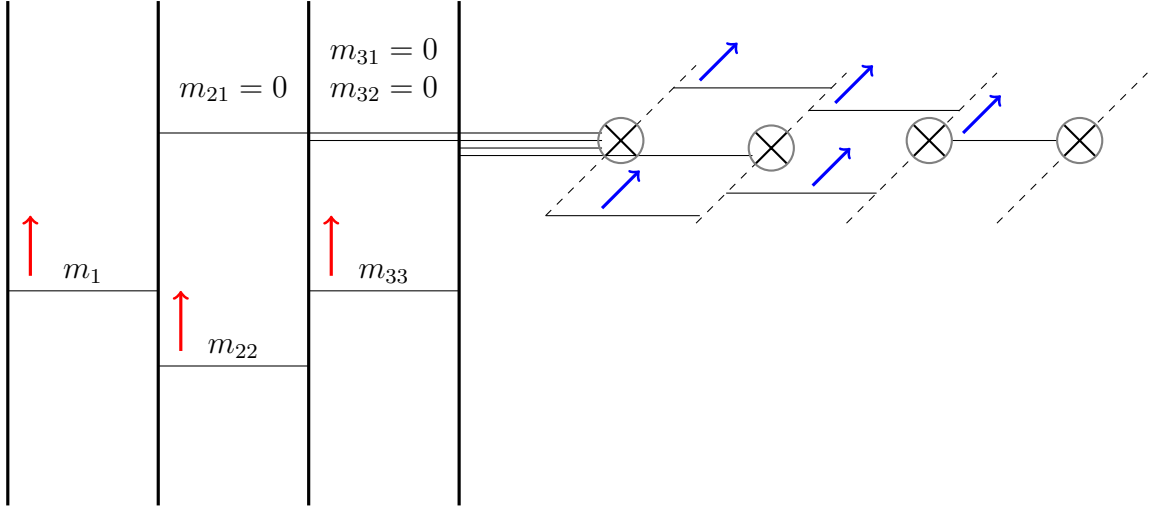
\begin{figure}[t]
\begin{center}
\begin{tikzpicture}
\node (u1u) at ( 0,4)  {};
\node (u1d) at ( 0,-3)  {}
edge [lpre] node[auto,swap] {} (u1u);
\node (u2u) at ( 2,4)  {};
\node (u2d) at ( 2,-3)  {}
edge [lpre] node[auto,swap] {} (u2u);
\node (u3u) at ( 4,4)  {};
\node (u3d) at ( 4,-3)  {}
edge [lpre] node[auto,swap] {} (u3u);
\node (u4u) at ( 6,4)  {};
\node (u4d) at ( 6,-3)  {}
edge [lpre] node[auto,swap] {} (u4u);

\draw (0,0)--(2,0);
\draw (2,2.1)--(7.9,2.1);
\draw (4,2)--(7.9,2);
\draw (6,1.9)--(7.9,1.9);
\draw (4,0)--(6,0);
\draw (2,-1)--(4,-1);
\draw (6,1.8)--(9.9,1.8);

\draw [dashed] (7.15,1)--(9.15,3);
\draw [dashed] (9.15,0.9)--(11.15,2.9);
\draw [dashed] (11.15,0.9)--(13.15,2.9);
\draw [dashed] (13.15,0.9)--(15.15,2.9);
\draw (8.85,2.7)--(10.95,2.7);
\draw (9.55,1.3)--(11.55,1.3);
\draw (10.65,2.4)--(12.65,2.4);
\draw (12.55,2)--(13.95,2);
\draw (7.15,1)--(9.2,1);

\node [below] at (1,0.5) {$m_1$};
\node [below] at (3,3) {$m_{21}=0$};
\node [below] at (3,-0.5) {$m_{22}$};
\node [below] at (5,3.5) {$m_{31}=0$};
\node [below] at (5,3) {$m_{32}=0$};
\node [below] at (5,0.5) {$m_{33}$};

\node [DfiveBig] (d51) at (8.15,2) {};
\node [DfiveBig] (d52) at (10.15,1.9) {};
\node [DfiveBig] (d53) at (14.25,2) {};
\node [DfiveBig] (d54) at (12.25,2) {};

\draw [->, very thick, blue] (9.2,2.8)--(9.7,3.3);
\draw [->, very thick, blue] (11,2.5)--(11.5,3);
\draw [->, very thick, blue] (12.7,2.1)--(13.2,2.6);
\draw [->, very thick, blue] (10.7,1.4)--(11.2,1.9);
\draw [->, very thick, blue] (7.9,1.1)--(8.4,1.6);
\draw [->, very thick, red] (0.3,0.2)--(0.3,1);
\draw [->, very thick, red] (4.3,0.2)--(4.3,1);
\draw [->, very thick, red] (2.3,-0.8)--(2.3,0);
\end{tikzpicture}
\caption{The brane picture for the $\rho=[3,1]$ mixed branch.}
\label{threeone}
\end{center}
\end{figure}

The restriction of the magnetic charges corresponding to $[3, 1]$ is 
\begin{equation}
\left(m_{21} = 0 \ \text{or} \ m_{22} = 0\right)\ \mbox{and}\ \left(m_{31}=m_{32}=0 \ \text{or} \ m_{32}=m_{33}=0\right),
\end{equation}
giving in total $4$ possible choices. We have to apply each one of them to equation \eqref{fullCoulombTSU4} and sum the four resulting sub-sums obtained. After performing such a restriction to the Hilbert series of the full Coulomb branch of the $T[SU(4)]$ theory, as explained in section \ref{sec:rule}, we find the following Hilbert series:
\begin{equation}
\begin{aligned}
HS(z,t)&=\sum_{k=0}^{9}[k,0,k]_zt^k+\mathcal{O}(t^{10}). \label{HS31}
\end{aligned}
\end{equation}
We also see the enhancement of the topological symmetry to $SU(4)$ since \eqref{HS31} is written by the characters of the $su(4)$ Lie algebra.

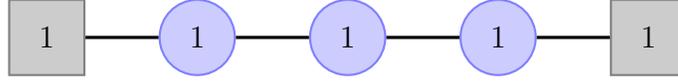
\begin{figure}[t]
\begin{center}
\begin{tikzpicture}
\node (m1) at ( -3,0) [global] {$1$};
\node (m2) at ( -1,0) [gauge] {$1$}
edge [lpre] node[auto,swap] {} (m1);
\node (m3) at ( 1,0) [gauge] {$1$}
edge [lpre] node[auto,swap] {} (m2);
\node (m4) at ( 3,0) [gauge] {$1$}
edge [lpre] node[auto,swap] {} (m3);
\node (f1) at ( 5,0) [global] {$1$}
edge [lpre] node[auto,swap] {} (m4);
\end{tikzpicture}
\caption{The quiver graph for the $T^{[3,1]}[SU(4)]$ theory.\label{31IR}}
\end{center}
\end{figure}

On the other hand, the IR theory at the infinitely large Higgs vev is given by the linear quiver of figure \ref{31IR}.
The dimension formula of this latter IR theory will read
\begin{equation}
\begin{aligned}
\Delta(n_1, n_2, n_3)&=\dfrac{1}{2}\left(|n_1|+|n_2-n_1|+|n_3-n_2|+|n_3|\right),
\end{aligned}
\end{equation}
where $n_1, n_2, n_3$ are the magnetic charges of the three $U(1)$'s.

The Hilbert series for the IR theory is given by
\begin{equation}
\begin{aligned}
HS(z,t)&=\sum_{n_1=-\infty}^{\infty}\sum_{n_2=-\infty}^{\infty}\sum_{n_3=-\infty}^{\infty}t^{\Delta(n_1,n_2,n_3)}z_1^{n_1}z_2^{n_2}z_3^{n_3}P_{U(1)}\left(n_1,t\right)P_{U(1)}\left(n_2,t\right)P_{U(1)}\left(n_3,t\right)
\end{aligned}
\end{equation}

By computing this explicitly we get 
\begin{equation}
\begin{aligned}
HS(z,t)&=\sum_{k=0}^{9}[k,0,k]_zt^k+\mathcal{O}(t^{10}),
\end{aligned}
\end{equation}
which precisely agrees with \eqref{HS31}. This matching has been checked up to order $12$ in $t$.

\subsubsection{The mixed branch $\rho=[3,2]$}

\begin{figure}[t]
\begin{center}
\begin{tikzpicture}
\node (m1) at ( -4,0) [gauge] {$1$};
\node (m2) at ( -2,0) [gauge] {$2$}
edge [lpre] node[auto,swap] {} (m1);
\node (m3) at ( 0,0) [gauge] {$3$}
edge [lpre] node[auto,swap] {} (m2);
\node (m4) at ( 2,0) [gauge] {$4$}
edge [lpre] node[auto,swap] {} (m3);
\node (m5) at ( 4,0) [global] {$5$}
edge [lpre] node[auto,swap] {} (m4);
\end{tikzpicture}
\caption{The quiver graph for $T[SU(5)]$. \label{32CB}}
\end{center}
\end{figure}
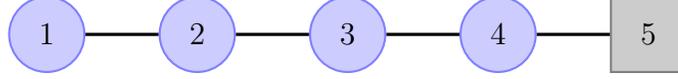

As a final example now we consider the mixed branch $\rho=[3,2]$ of the $T[SU(5)]$ theory. The quiver description of the $T[SU(5)]$ is given by figure \ref{32CB}.
We denote the magnetic charge of the $U(1)$ by $m_{11}$, the magnetic charges of the $U(2)$ by $m_{21}, m_{22}$, the magnetic charges of the $U(3)$ by $m_{31}, m_{32}, m_{33}$ and the magnetic charges of the $U(5)$ by $m_{41}, m_{42}, m_{43}, m_{44}$. The dimension formula for the full Coulomb branch of $T[SU(5)]$ is given by
\begin{equation}
\begin{aligned}
\Delta(\vec{m})=& -|m_{21}-m_{22}|-|m_{31}-m_{32}|-|m_{31}-m_{33}|-|m_{32}-m_{33}|-|m_{41}-m_{42}|\\
&-|m_{41}-m_{43}|-|m_{41}-m_{44}|-|m_{42}-m_{43}|-|m_{42}-m_{44}|-|m_{43}-m_{44}|\\
&+\dfrac{1}{2}\left(|m_{21}-m_{11}|+|m_{22}-m_{11}|+|m_{31}-m_{21}|+|m_{31}-m_{22}|+|m_{32}-m_{21}|\right.\\
&+|m_{32}-m_{22}|+|m_{33}-m_{21}|+|m_{33}-m_{22}|+|m_{41}-m_{31}|+|m_{41}-m_{32}|\\
&+|m_{41}-m_{33}|+|m_{42}-m_{31}|+|m_{42}-m_{32}|+|m_{42}-m_{33}|+|m_{43}-m_{31}|\\
&+|m_{43}-m_{32}|+|m_{43}-m_{33}|+|m_{44}-m_{31}|+|m_{44}-m_{32}|+|m_{44}-m_{33}|\\
&\left.+5|m_{41}|+5|m_{42}|+5|m_{43}|+5|m_{44}|\right),
\end{aligned}
\end{equation}
where $\vec{m}=\left(m_{11}, m_{21}, m_{22}, m_{31}, m_{32}, m_{33}, m_{41}, m_{42}, m_{43}, m_{44}\right)$ satisfying $m_{21} \geq m_{22}$, $m_{31} \geq m_{32} \geq m_{33}$ and $m_{41} \geq m_{42} \geq m_{43} \geq m_{44}$. 

The Hilbert Series for the full Coulomb branch of this theory is
\begin{equation}
\begin{aligned}
HS(z_,t)&:=\sum_{m_1=-\infty}^{\infty}\sum_{m_{21}\geq m_{22}}\sum_{m_{31}\geq m_{32}\geq m_{33}}\sum_{m_{41}\geq m_{42}\geq m_{43}\geq m_{44}}t^{\Delta(m_1,m_{21},m_{22},m_{31},m_{32},m_{33},m_{41},m_{42},m_{43},m_{44})}\\
&\cdot P_{U(1)}(m_1,t) P_{U(2)}(m_{21},m_{22},t) P_{U(3)}(m_{31},m_{32},m_{33},t)P_{U(4)}(m_{41},m_{42},m_{43},m_{44},t)\\
&\cdot z_1^{m_1}z_2^{\left(m_{21}+m_{22}\right)}z_3^{\left(m_{31}+m_{32}+m_{33}\right)}z_4^{\left(m_{41}+m_{42}+m_{43}+m_{44}\right)}.
\label{HSC32}
\end{aligned}
\end{equation}

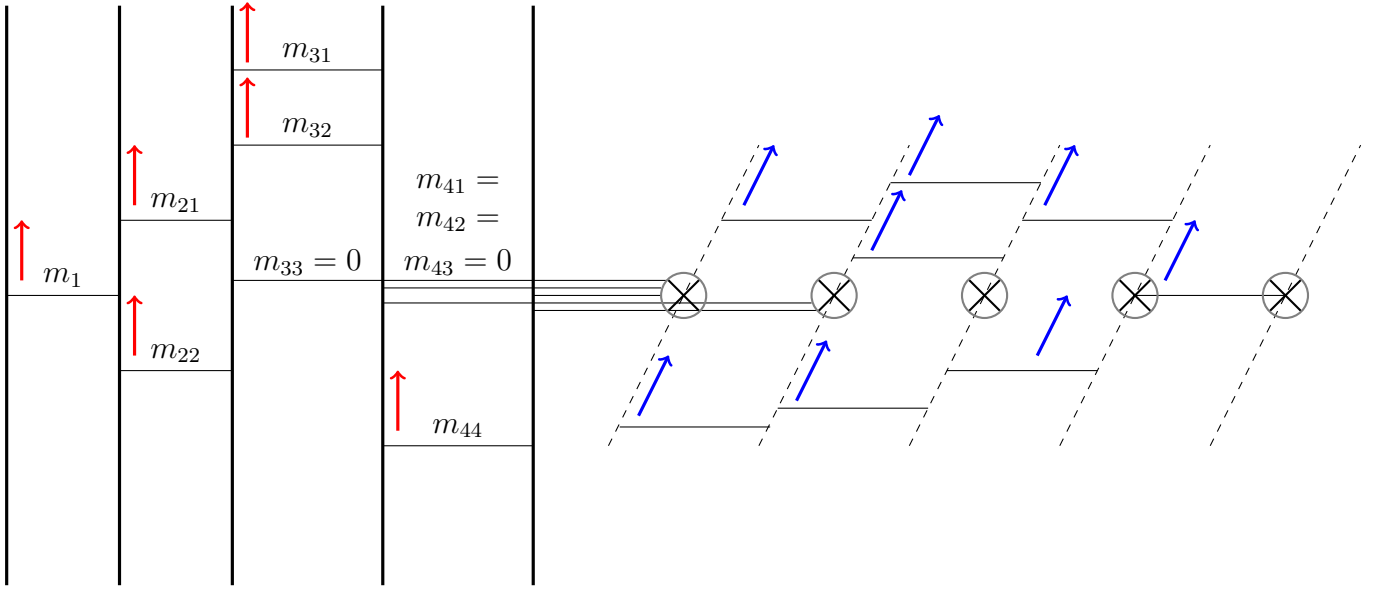
\begin{figure}[t]
\begin{center}
\begin{tikzpicture}
\node (u3u) at ( -6,4)  {};
\node (u3d) at ( -6,-4)  {}
edge [lpre] node[auto,swap] {} (u3u);
\node (u5u) at ( -4.5,4)  {};
\node (u5d) at ( -4.5,-4)  {}
edge [lpre] node[auto,swap] {} (u5u);
\node (u4u) at ( -3,4)  {};
\node (u4d) at ( -3,-4)  {}
edge [lpre] node[auto,swap] {} (u4u);
\node (u1u) at ( -1,4)  {};
\node (u1d) at ( -1,-4)  {}
edge [lpre] node[auto,swap] {} (u1u);
\node (u2u) at ( 1,4)  {};
\node (u2d) at ( 1,-4)  {}
edge [lpre] node[auto,swap] {} (u2u);

\draw (-6,0)--(-4.5,0);

\draw (-3,0.2)--(2.8,0.2);
\draw (-1,0.1)--(2.7,0.1);
\draw (1,0)--(2.7,0);
\draw (-1,-0.1)--(4.7,-0.1);
\draw (1,-0.2)--(4.8,-0.2);

\draw(-4.5,1)--(-3,1);
\draw(-4.5,-1)--(-3,-1);

\draw(-3,3)--(-1,3);
\draw(-3,2)--(-1,2);

\draw(-1,-2)--(1,-2);

\node [below] at (-5.25,0.5) {$m_1$};
\node [below] at (-3.75,1.5) {$m_{21}$};
\node [below] at (-3.75,-0.5) {$m_{22}$};
\node [below] at (-2,3.5) {$m_{31}$};
\node [below] at (-2,2.5) {$m_{32}$};
\node [below] at (-2,0.75) {$m_{33}=0$};
\node [below] at (0,1.75) {$m_{41}=$};
\node [below] at (0,1.25) {$m_{42}=$};
\node [below] at (0,0.75) {$m_{43}=0$};
\node [below] at (0,-1.5) {$m_{44}$};

\draw [dashed] (2,-2)--(4,2);
\draw [dashed] (4,-2)--(6,2);
\draw [dashed] (6,-2)--(8,2);
\draw [dashed] (8,-2)--(10,2);
\draw [dashed] (10,-2)--(12,2);

\draw (3.5,1)--(5.5,1);
\draw (2.15,-1.75)--(4.15,-1.75);

\draw (5.75,1.5)--(7.75,1.5);
\draw (4.25,-1.5)--(6.25,-1.5);
\draw (5.25,0.5)--(7.25,0.5);

\draw (7.5,1)--(9.5,1);
\draw (6.5,-1)--(8.5,-1);
\draw (9,0)--(11,0);

\node [DfiveBig] (d53) at (3,0) {};
\node [DfiveBig] (d53) at (5,0) {};
\node [DfiveBig] (d51) at (7,0) {};
\node [DfiveBig] (d51) at (9,0) {};
\node [DfiveBig] (d52) at (11,0) {};

\draw [->, very thick, red] (-5.8,0.2)--(-5.8,1);
\draw [->, very thick, red] (-4.3,1.2)--(-4.3,2);
\draw [->, very thick, red] (-4.3,-0.8)--(-4.3,0);
\draw [->, very thick, red] (-2.8,2.1)--(-2.8,2.9);
\draw [->, very thick, red] (-2.8,3.1)--(-2.8,3.9);
\draw [->, very thick, red] (-0.8,-1.8)--(-0.8,-1);

\draw [->, very thick, blue] (3.8,1.2)--(4.2,2);
\draw [->, very thick, blue] (2.4,-1.6)--(2.8,-0.8);
\draw [->, very thick, blue] (6,1.6)--(6.4,2.4);
\draw [->, very thick, blue] (5.5,0.6)--(5.9,1.4);
\draw [->, very thick, blue] (4.5,-1.4)--(4.9,-0.6);
\draw [->, very thick, blue] (7.7,-0.8)--(8.1,0);
\draw [->, very thick, blue] (7.8,1.2)--(8.2,2);
\draw [->, very thick, blue] (9.4,0.2)--(9.8,1);

\end{tikzpicture}
\caption{The brane picture for the $\rho=[3,2]$ mixed branch, for the subcase in which $m_{41}=m_{42}=m_{43}=0 \geq m_{44}$ and $m_{31}\geq m_{32} > m_{33}=0$.}
\label{MegaPicture}
\end{center}
\end{figure}

By going to the mixed branch we wish to analyze, we have the brane picture in figure \ref{MegaPicture}. We see that by using the first rule we have to set to zero $3$ magnetic charges of the $4$th cell, and $1$ magnetic charge of the $3$rd cell. Again, there are different ways to do so: in the $4$th cell we can have $m_{41}=m_{42}=m_{43}=0$ or $m_{42}=m_{43}=m_{44}=0$. For any of these two cases, we have three choices in the $3$rd cell, namely $m_{31}=0$, $m_{32}=0$ or $m_{33}=0$. In total, we find six different sub-cases into which equation (\ref{HSC32}) splits, and we must sum over all of them.

By performing the summation over all these subcases we find the following Hilbert series.
\begin{equation}
\begin{aligned}
HS(z,t)&=1 + [1,0,0,1]_z t + \left([2,0,0,2]_z+[0,1,1,0]_z\right) t^2 \\
&+ \left([3,0,0,3]_z+[1,1,1,1]_z\right) t^3 + \left([4,0,0,4]_z+[2,1,1,2]_z+[0,2,2,0]_z\right) t^4 \\
&+\left([5,0,0,5]_z+[3,1,1,3]_z+[1,2,2,1]_z\right) t^5 +\\
&+\left([6,0,0,6]_z+[4,1,1,4]_z+[2,2,2,2]_z+[0,3,3,0]_z\right) t^6 +\\
&+\left([7,0,0,7]_z+[5,1,1,5]_z+[3,2,2,3]_z+[1,3,3,1]_z\right) t^7 +\\
&+\left([8,0,0,8]_z+[6,1,1,6]_z+[4,2,2,4]_z+[2,3,3,2]_z+[0,4,4,0]_z\right) t^8 +\\
&+\left([9,0,0,9]_z+[7,1,1,7]_z+[5,2,2,5]_z+[3,3,3,3]_z+[1,4,4,1]_z\right) t^9 +\mathcal{O}(t^{10}).
\end{aligned}
\label{MegaHSUV}
\end{equation}
The Hilbert series is written by the characters of the $su(5)$ Lie algebra, and this implies that the topological symmetry is enhanced to $SU(5)$.

\begin{figure}[t]
\begin{center}
\begin{tikzpicture}
\node (m1) at ( -3,0) [gauge] {$1$};
\node (m2) at ( -1,0) [gauge] {$2$}
edge [lpre] node[auto,swap] {} (m1);
\node (m3) at ( 1,0) [gauge] {$2$}
edge [lpre] node[auto,swap] {} (m2);
\node (m4) at ( 3,0) [gauge] {$1$}
edge [lpre] node[auto,swap] {} (m3);
\node (f1) at ( -1,2) [global] {$1$}
edge [lpre] node[auto,swap] {} (m2);
\node (f2) at ( 1,2) [global] {$1$}
edge [lpre] node[auto,swap] {} (m3);
\end{tikzpicture}
\caption{The Quiver graph for the $T^{[3,2]}[SU(5)]$ theory.}
\label{MegaIRHW}
\end{center}
\end{figure}
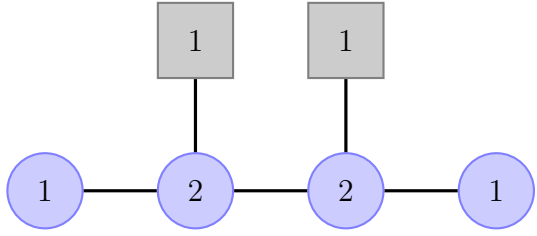

Now we would like to check this result by using the same procedure of the other examples. After going to the IR by giving infinitely large vev to the hypermultiplets and performing some Hanany-Witten transitions, we find the quiver theory of figure \ref{MegaIRHW}.
For this linear quiver theory, the dimension formula is
\begin{equation}
\begin{aligned}
\Delta(m_1, m_{21}, m_{22}, n_{21}, n_{22}, n_{1})&=-|m_{22}-m_{21}|-|n_{22}-n_{21}|+\dfrac{1}{2}(|m_{21}-m_{1}|+|m_{22}-m_{1}|\\
&+|n_{21}-m_{21}|+|n_{21}-m_{22}|+|n_{22}-m_{21}|+|n_{22}-m_{22}|\\
&+|n_{1}-n_{21}|+|n_{1}-n_{22}|+|m_{21}|+|m_{22}|+|n_{21}|+|n_{22}|).
\end{aligned}
\end{equation}
where we assign magnetic charges as follows: $m_1$ is the magnetic charge for the leftmost $U(1)$, $m_{21}$ and $m_{22}$ (resp. $n_{21}$ and $n_{22}$) are the magnetic charges for the leftmost (resp. rightmost) $U(2)$ group, and $n_1$ for the rightmost $U(1)$.
The Hilbert series for the IR theory is given by
\begin{equation}
\begin{aligned}
HS(z,t)&=\sum_{m_1=-\infty}^{\infty}\sum_{n_1=-\infty}^{\infty}\sum_{n_{21}\geq n_{22}}\sum_{m_{21}\geq m_{22}}t^{\Delta(m_1,m_{21},m_{22},n_{21},n_{22},n_1)}z_1^{m_1}z_2^{(m_{21}+m_{22})}z_3^{(n_{21}+n_{22})}z_4^{n_1}\\
&\cdot
P_{U(1)}\left(n_1,t\right)P_{U(1)}\left(n_2,t\right)P_{U(2)}\left(n_{21},n_{22},t\right)P_{U(2)}\left(m_{21},m_{22},t\right)
\end{aligned}
\end{equation}

By explicitly computing the refined Hilbert series we get
\begin{equation}
\begin{aligned}
HS(z,t)&=1 + [1,0,0,1]_z t + \left([2,0,0,2]_z+[0,1,1,0]_z\right) t^2 \\
&+ \left([3,0,0,3]_z+[1,1,1,1]_z\right) t^3 + \left([4,0,0,4]_z+[2,1,1,2]_z+[0,2,2,0]_z\right) t^4 \\
&+\left([5,0,0,5]_z+[3,1,1,3]_z+[1,2,2,1]_z\right) t^5 +\\
&+\left([6,0,0,6]_z+[4,1,1,4]_z+[2,2,2,2]_z+[0,3,3,0]_z\right) t^6 +\\
&+\left([7,0,0,7]_z+[5,1,1,5]_z+[3,2,2,3]_z+[1,3,3,1]_z\right) t^7 +\\
&+\left([8,0,0,8]_z+[6,1,1,6]_z+[4,2,2,4]_z+[2,3,3,2]_z+[0,4,4,0]_z\right) t^8 +\\
&+\left([9,0,0,9]_z+[7,1,1,7]_z+[5,2,2,5]_z+[3,3,3,3]_z+[1,4,4,1]_z\right) t^9 +\mathcal{O}(t^{10}).
\end{aligned}
\end{equation}

and we see this is in perfect agreement with the Hilbert series found directly by the restriction rule in equation \eqref{MegaHSUV}. This matching has been checked up to order $9$ in $t$.

\section{Higgs Branch Examples}
\label{sec:Higgs}

The Hilbert series for the Higgs branch part of any mixed branch of the $T[SU(N)]$ theory can be computed by going to the IR by decoupling all the Coulomb branch moduli which are not frozen, as explained in section \ref{sec:HSHiggspart}, and by using the method described in section \ref{sec:HSHiggs}. In section \ref{sec:rule}, we use yet another way to compute the Hilbert series of the Higgs branch by using the restriction rule as well as the 3d mirror symmetry.  In this section we will compute explicitly the Hilbert series for the Higgs branch part of all the mixed branches considered in section \ref{sec:Coulombexs} by using the two methods. We will in fact find the complete agreement between the two results which give a nice check for the restriction rule as well as the 3d mirror symmetry. 

\subsection{The mixed branch $\rho=[2,2]$}

In this case, we are interested in the Higgs branch part of the $[2,2]$ branch. The brane picture for this mixed branch is given already in figure \ref{brane1234A}. We first compute the Hilbert series of the Higgs branch factor $\mathcal{H}_{[2,2]}$ by using the method described in \ref{sec:HSHiggspart}. For that we make use of the IR theory of \eqref{IRHiggs}. 

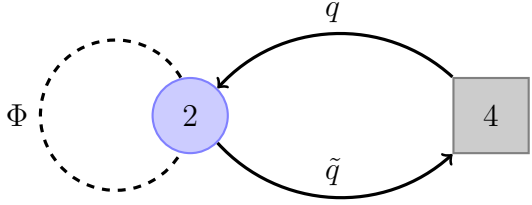
\begin{figure}[t]
\begin{center}
\begin{tikzpicture}
\node (m1) at ( -2,0) [gauge] {$2$};
\cercle{-3,0}{1cm}{30}{300};
\node (m2) at ( 2,0) [global] {$4$}
edge [<-, very thick, bend left=45] node[auto, swap] {$\tilde{q}$} (m1)
edge [->, very thick, bend right=45] node[auto, swap] {$q$} (m1);
\node (m2) at ( -4.3,0) {$\Phi$};
\end{tikzpicture}
\caption{The quiver graph for the $U(2)$ gauge theory with $4$ flavors. $q, \tilde{q}, \Phi$ are 3d $\mathcal{N}=2$ chiral multiplets and $q, \tilde{q}$ form a 3d $\mathcal{N}=4$ hypermultiplet. }
\label{(2)[4]}
\end{center}
\end{figure}

By decoupling all the unfrozen Coulomb branch moduli (i.e. sending to infinity the mobile color D3-branes) we see that we are left only with $2$ frozen color D3-branes in the $3$rd cell, and $4$ flavor D3-branes.
The IR theory is therefore given by $U(2)$ with $4$ flavors.
In figure \ref{(2)[4]} we describe the matter in this IR theory, by using a 3d $\mathcal{N}=2$ quiver notation, in which one 3d $\mathcal{N}=4$ hypermultiplet is split in two 3d $\mathcal{N}=2$ chiral multiplets $\left(q,\tilde{q}\right)$, and we explicitly write a 3d $\mathcal{N}=2$ chiral multiplet $\Phi$ in the adjoint representation of the $U(2)$, which lies inside a 3d $\mathcal{N}=4$ vector multiplet.
From the quiver, we can read the charge assignment of all the different chiral multiplets, and we then associate fugacities to the gauge and flavor symmetry groups, according to table \ref{table(2)[4]}. In our notation, if an arrow is pointing toward a group $G$, then a chiral multiplet associated to the arrow is in the fundamental representation under the symmetry group $G$. 

\begin{table}[t]
\centering
\begin{tabular}{cccc}
\cline{2-3}
 & \multicolumn{2}{|c|}{$U(2)_g$} &  \\ 
\cline{2-4} 
\multicolumn{1}{c|}{} & \multicolumn{1}{|c|}{$U(1)_g$} & \multicolumn{1}{|c|}{$SU(2)_g$} & \multicolumn{1}{|c|}{$SU(4)_f$}\\
\cline{1-4}
\multicolumn{1}{|c|}{$q$}&\multicolumn{1}{|c|}{$w_{1}$}&\multicolumn{1}{|c|}{$[1]_{w_2}$}&\multicolumn{1}{|c|}{$[0,0,1]_{x_1,x_2,x_3}$}\\ 
\hline
\multicolumn{1}{|c|}{$\tilde{q}$}&\multicolumn{1}{|c|}{$w_1^{-1}$}&\multicolumn{1}{|c|}{$[1]_{w_2}$}&\multicolumn{1}{|c|}{$[1,0,0]_{x_1,x_2,x_3}$}\\ 
\hline
\multicolumn{1}{|c|}{$\Phi$}&\multicolumn{1}{|c|}{$1$}&\multicolumn{1}{|c|}{$[2]_{w_2}$}&\multicolumn{1}{|c|}{$1$}\\ 
\hline
\end{tabular}
\caption{The charge assignment of the fields of the $\tilde{T}^{[2,2]}[SU(4)]$ theory, as seen from the quiver in figure \ref{(2)[4]}. \label{table(2)[4]}}
\end{table}

The quiver in figure \ref{(2)[4]} also has the following superpotential in terms of the 3d $\mathcal{N}=2$ notation,
\begin{equation}
W=\tr(q^i\Phi_{i\bar{j}}\tilde{q}^{\bar{j}}),
\end{equation}
where $i$ (resp. $\bar{j}$) is an index of the fundamental (resp. anti-fundamental) representation of the $U(2)$, and the trace is performed on the flavor indices.
By deriving the F-term equations from this superpotential by taking a derivative with respect to $\Phi_{i\bar{j}}$, we notice that on the Higgs branch there is one relation of order $2$ in $\tilde{t}$, and carrying both an index $i$ and an index $\bar{j}$. This splits into two independent equations: one in the adjoint and the other in the trivial representation of the $U(2)$. The other F-term conditions are automatically satisfied since the vevs for $\Phi_{i\bar{j}}$ are zero. Out of this information we can derive the F-term prefactor described in section \ref{sec:HSHiggs} as
\begin{equation}
\mbox{Pfc}(w_2,\tilde{t})=\left(\PE\left[[2]_{w_2}\tilde{t}^2+\tilde{t}^2\right]\right)^{-1},
\end{equation}
which we will need to multiply to the integrand of formula of \eqref{HSHiggs.comp}, as explained in section \ref{sec:HSHiggs}.

Therefore the Hilbert series is given by
\begin{equation}
\begin{aligned}
HS\left(\tilde{t},x\right)&=\int d\mu_{U(2)}\ \ \mbox{Pfc}(w_2,\tilde{t})\cdot \PE\left[w_{1}[1]_{w_2}[0,0,1]_{x}\tilde{t}+w_{1}^{-1}[1]_{w_2}[1,0,0]_{x}\tilde{t}\right],
\end{aligned}
\end{equation}
where we recall that the Haar measure for a $U(2)$ gauge group is given by 
\begin{equation}
\int d\mu_{U(2)}=\dfrac{1}{(2\pi i)^2}\oint_{|w_1|=1}\oint_{|w_2|=1} \dfrac{dw_1}{w_{1}}\dfrac{dw_2}{w_{2}}(1-w_2^2).
\end{equation}
By performing explicitly this residue computation we get the following result,
\begin{equation}
HS\left(x,t\right)=1+[1,0,1]_{x}t+\left([2,0,2]_x+[0,2,0]_x\right)t^2 + \mathcal{O}(t^3), \label{HS22fromHiggs}
\end{equation}
where we write the expression in terms of $t = \tilde{t}^{2}$.

\subsubsection*{Matching with the dual Coulomb branch part of $[2,2]$}

We then move on to the other method of the computation in section \ref{sec:rule} where we use the restriction rule and the 3d mirror symmetry. The dual partition to $[2, 2]$ is in fact $[2, 2]$. Therefore we can reuse the result of \eqref{HS22}.  
By keeping track of the fugacities $z_i$ for the topological symmetry the result was
\begin{equation}
\begin{aligned}
HS(t,z)&=1+[1,0,1]_{z}t+\left([2,0,2]_{z}+[0,2,0]_{z}\right)t^2 + \mathcal{O}(t^3),
\end{aligned}
\end{equation}
which exactly coincides with \eqref{HS22fromHiggs}. This matching has been checked up to order $7$ in $t$.  Note that $z$ is related to $x$ by performing a redefinition of the Cartan generators as explained in \eqref{basisz} and \eqref{basisx}. In this case, the relations are
\begin{equation}
z_1\mapsto x_1^2x_2^{-1}, \quad z_2\mapsto x_1^{-1}x_2^{2},
\end{equation}
since the Cartan matrix of the $su(3)$ Lie algebra is given by  
\begin{equation}
M_{A_2}=\left(\begin{array}{cc}
2 & -1 \\ 
-1 & 2
\end{array} \right).
\end{equation}

\subsection{The mixed branch $\rho=[3,1]$}
\begin{figure}[t]
\begin{center}
\begin{tikzpicture}
\node (m1) at ( -2,0) [gauge] {$1$};
\cercle{-3,0}{1cm}{30}{300};
\node (m2) at ( 2,0) [gauge] {$2$}
edge [<-, very thick, bend left=45] node[auto, swap] {$\tilde{q_1}$} (m1)
edge [->, very thick, bend right=45] node[auto, swap] {$q_1$} (m1);
\node (m3) at ( 6,0) [global] {$4$}
edge [<-, very thick, bend left=45] node[auto, swap] {$\tilde{q_2}$} (m2)
edge [->, very thick, bend right=45] node[auto, swap] {$q_2$} (m2);
\node (m2) at ( -4.3,0) {$\Phi_1$};
\draw[very thick, dashed, -latex] (2,1) [partial ellipse=-30:220:0.4cm and 0.8cm];
\node at ( 2,2.2) {$\Phi_2$};
\end{tikzpicture}
\caption{The quiver graph for the $\tilde{T}^{[2,2]}[SU(4)]$ theory.}
\label{(1)(2)[4]}
\end{center}
\end{figure}
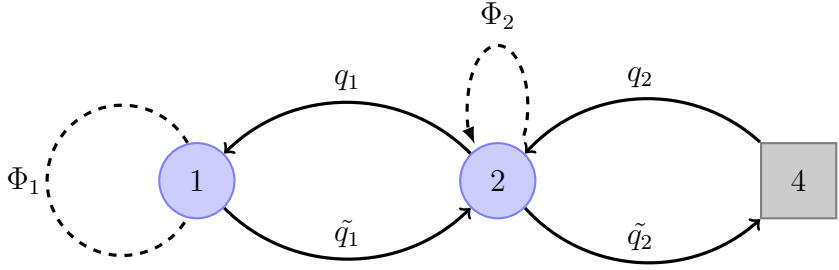

The next example is the Higgs branch part of the mixed branch $[3,1]$. The brane picture for this mixed branch is given in figure \ref{threeone}. We first compute the Hilbert series by the method in section \ref{sec:HSHiggspart} and hence we send to infinity all the unfrozen color D3-branes. Doing this, we end up with a brane diagram yielding a theory given by the quiver depicted in figure \ref{(1)(2)[4]}. Our claim is that the Hilbert series for the full Higgs branch of this theory is the same as the Higgs branch part of the mixed branch of $[3,1]$.
Therefore we compute the Hilbert series, using the techniques in section \ref{sec:HSHiggs}.

From figure \ref{(1)(2)[4]} we can read what are the matter fields. We assign gauge fugacities $w_1$ to the $U(1)_1$ factor of the gauge group, and $w_2, w_3$ to the $U(2)_2$ factor. In particular $w_2$ will be the fugacity for the overall $U(1)_2\hookrightarrow U(2)_2$ and $w_3$ for $SU(2)_2\hookrightarrow U(2)_2$. We also assign fugacities $x_1, x_2, x_3$ to the flavor $SU(4)$ group.
We summarize the matter fields and their charges in table \ref{table(1)(2)[4]}. 
\begin{table}[t]
\centering
\begin{tabular}{ccccc}
\cline{3-4}
& & \multicolumn{2}{|c|}{$U(2)_2$} &  \\ 
\cline{2-5} 
 & \multicolumn{1}{|c|}{$U(1)_1$} & \multicolumn{1}{|c|}{$U(1)_2$} & \multicolumn{1}{|c|}{$SU(2)_2$} & \multicolumn{1}{|c|}{$SU(4)_f$}\\
\cline{1-5}
\multicolumn{1}{|c|}{$q_1$}&\multicolumn{1}{|c|}{$w_1^{1}$} &\multicolumn{1}{|c|}{$w_2^{-1}$}&\multicolumn{1}{|c|}{$[1]_{w_2}$}&\multicolumn{1}{|c|}{$1$}\\ 
\hline
\multicolumn{1}{|c|}{$\tilde{q}_1$}& \multicolumn{1}{|c|}{$w_1^{-1}$} &\multicolumn{1}{|c|}{$w_2^{1}$}&\multicolumn{1}{|c|}{$[1]_{w_2}$}&\multicolumn{1}{|c|}{$1$}\\ 
\hline
\multicolumn{1}{|c|}{$q_2$}&\multicolumn{1}{|c|}{$1$} &\multicolumn{1}{|c|}{$w_2^{1}$}&\multicolumn{1}{|c|}{$[1]_{w_2}$}&\multicolumn{1}{|c|}{$[0,0,1]_{x_1,x_2,x_3}$}\\ 
\hline
\multicolumn{1}{|c|}{$\tilde{q}_2$}& \multicolumn{1}{|c|}{$1$} &\multicolumn{1}{|c|}{$w_2^{-1}$}&\multicolumn{1}{|c|}{$[1]_{w_2}$}&\multicolumn{1}{|c|}{$[1,0,0]_{x_1,x_2,x_3}$}\\ 
\hline
\multicolumn{1}{|c|}{$\Phi_1$}& $1$ &\multicolumn{1}{|c|}{$1$}&\multicolumn{1}{|c|}{$1$}&\multicolumn{1}{|c|}{$1$}\\ 
\hline
\multicolumn{1}{|c|}{$\Phi_2$}& $1$ &\multicolumn{1}{|c|}{$1$}&\multicolumn{1}{|c|}{$[2]_{w_3}$}&\multicolumn{1}{|c|}{$1$}\\ 
\hline
\end{tabular}
\caption{The charge assignment of the fields in the $\tilde{T}^{[3,1]}[SU(4)]$ theory, as seen from the quiver in figure \ref{(1)(2)[4]}.\label{table(1)(2)[4]}}
\end{table} 

From the quiver in figure \ref{(1)(2)[4]}, we can also read off the superpotential that in this case leads to F-term constraints generating a prefactor
\begin{equation}
\mbox{Pfc}(w_3,\tilde{t})=\left(\PE\left[[2]_{w_3}\tilde{t}^2+2\tilde{t}^2\right]\right)^{-1}.
\end{equation}

Therefore the Hilbert series is given by
\begin{equation}
\begin{aligned}
HS\left(x,\tilde{t}\right)&=\int d\mu_{U(1)\times U(2)}\ \ \mbox{Pfc}(w_3,\tilde{t})\ \cdot\\
&\cdot \PE\left[w_{1}w_2^{-1}[1]_{w_3}\tilde{t}+w_{1}^{-1}w_2[1]_{w_3}\tilde{t}+w_2[1]_{w_3}[0,0,1]_x\tilde{t}+w_2^{-1}[1]_{w_3}[1,0,0]_x\tilde{t}\right],
\end{aligned}
\end{equation}
where we recall that the Haar measure for a $U(1)\times U(2)$ gauge group is given by 
\begin{equation}
\int d\mu_{U(1)\times U(2)}=\dfrac{1}{(2\pi i)^3}\oint_{|w_1|=1}\oint_{|w_2|=1}\oint_{|w_3|=1} \dfrac{dw_1}{w_{1}}\dfrac{dw_2}{w_{2}}\dfrac{dw_3}{w_{3}}(1-w_3^2).
\end{equation}
By performing explicitly this residue computation we get to the following result:
\begin{equation}
HS(x,t)=1+[1,0,1]_xt+\left([2,0,2]_x+[0,2,0]_x+[1,0,1]_x\right)t^2+\mathcal{O}(t^3),\label{HS31fromHiggs}
\end{equation}
where we again used $t = \tilde{t}^2$.

\subsubsection*{Matching with the dual Coulomb branch part of $[2,1,1]$}

We then move on the the mirror computation in section \ref{sec:rule}. The dual to the partition $[3,1]$ is $[2,1,1]$. Hence we compute the Hilbert series of $\mathcal{C}_{[2,1,1]}$ by the restriction rule. We do not repeat the process of the computation and quote the result 
\begin{equation}
HS(t,z)=1+[1,0,1]_zt+\left([2,0,2]_z+[0,2,0]_z+[1,0,1]_z\right)t^2+\mathcal{O}(t^3), 
\end{equation}
which completely agrees with \eqref{HS31fromHiggs}. This matching has been checked up to order $6$ in $t$. Here $z$ is related to $x$ by performing a redefinition of the Cartan generators and it is given by 
\begin{equation}
z_1\mapsto x_1^2x_2^{-1}, \quad z_2\mapsto x_1^{-1}x_2^{2}x_3^{-1} \quad z_3\mapsto x_2^{-1}x_3^{2},
\end{equation}
from the Cartan matrix of the $su(3)$,
\begin{equation}
M_{A_2}=\left(\begin{array}{ccc}
2 & -1 & 0 \\ 
-1 & 2 & -1 \\ 
0 & -1 & 2 
\end{array} \right).
\end{equation}

\subsection{The mixed branch $\rho=[3,2]$}
\begin{figure}[t]
\begin{center}
\begin{tikzpicture}
\node (m1) at ( -2,0) [gauge] {$1$};
\cercle{-3,0}{1cm}{30}{300};
\node (m2) at ( 2,0) [gauge] {$3$}
edge [<-, very thick, bend left=45] node[auto, swap] {$\tilde{q_1}$} (m1)
edge [->, very thick, bend right=45] node[auto, swap] {$q_1$} (m1);
\node (m3) at ( 6,0) [global] {$5$}
edge [<-, very thick, bend left=45] node[auto, swap] {$\tilde{q_2}$} (m2)
edge [->, very thick, bend right=45] node[auto, swap] {$q_2$} (m2);
\node (m2) at ( -4.3,0) {$\Phi_1$};
\draw[very thick, dashed, -latex] (2,1) [partial ellipse=-30:220:0.4cm and 0.8cm];
\node at ( 2,2.2) {$\Phi_2$};
\end{tikzpicture}
\caption{The quiver graph for the $\tilde{T}^{[3,2]}[SU(5)]$ theory.}
\label{(1)(3)[5]}
\end{center}
\end{figure}
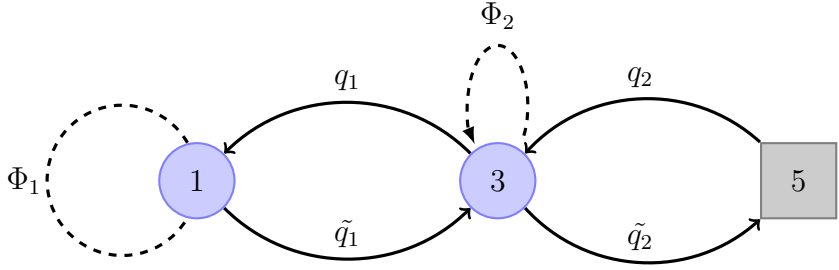

In this last example we are interested in the Higgs branch part of the mixed branch $[3,2]$. For doing the computation in section \ref{sec:HSHiggspart}, we use the quiver diagram of the $\tilde{T}^{[3,2]}[SU(5)]$ theory given in figure \ref{(1)(3)[5]} by using the general result \eqref{IRHiggs}. 

From the quiver, we can read the matter fields and their charges under the global and gauge symmetry groups. This is reported in table \ref{table(1)(3)[5]}. In particular we assign a fugacity $w_1$ to the $U(1)$ factor of the gauge group, fugacities $w_2, w_3$ and $w_4$ to the $U(3)$ gauge group, and fugacities $x_1, x_2, x_3, x_4$ to the $SU(5)$ flavor symmetry.

\begin{table}[t]
\centering
\begin{tabular}{ccccc}
\cline{3-4}
& & \multicolumn{2}{|c|}{$U(3)_2$} &  \\ 
\cline{2-5} 
 & \multicolumn{1}{|c|}{$U(1)_1$} & \multicolumn{1}{|c|}{$U(1)_2$} & \multicolumn{1}{|c|}{$SU(3)_2$} & \multicolumn{1}{|c|}{$SU(5)_f$}\\
\cline{1-5}
\multicolumn{1}{|c|}{$q_1$}&\multicolumn{1}{|c|}{$w_1^{1}$} &\multicolumn{1}{|c|}{$w_2^{-1}$}&\multicolumn{1}{|c|}{$[1,0]_{w_2, w_3}$}&\multicolumn{1}{|c|}{$1$}\\ 
\hline
\multicolumn{1}{|c|}{$\tilde{q}_1$}& \multicolumn{1}{|c|}{$w_1^{-1}$} &\multicolumn{1}{|c|}{$w_2^{1}$}&\multicolumn{1}{|c|}{$[0,1]_{w_2, w_3}$}&\multicolumn{1}{|c|}{$1$}\\ 
\hline
\multicolumn{1}{|c|}{$q_2$}&\multicolumn{1}{|c|}{$1$} &\multicolumn{1}{|c|}{$w_2^{1}$}&\multicolumn{1}{|c|}{$[0,1]_{w_2, w_3}$}&\multicolumn{1}{|c|}{$[0,0,0,1]_{x_1,x_2,x_3,x_4}$}\\ 
\hline
\multicolumn{1}{|c|}{$\tilde{q}_2$}& \multicolumn{1}{|c|}{$1$} &\multicolumn{1}{|c|}{$w_2^{-1}$}&\multicolumn{1}{|c|}{$[1,0]_{w_2, w_3}$}&\multicolumn{1}{|c|}{$[1,0,0,0]_{x_1,x_2,x_3,x_4}$}\\ 
\hline
\multicolumn{1}{|c|}{$\Phi_1$}& $1$ &\multicolumn{1}{|c|}{$1$}&\multicolumn{1}{|c|}{$1$}&\multicolumn{1}{|c|}{$1$}\\ 
\hline
\multicolumn{1}{|c|}{$\Phi_2$}& $1$ &\multicolumn{1}{|c|}{$1$}&\multicolumn{1}{|c|}{$[1,1]_{w_3,w_4}$}&\multicolumn{1}{|c|}{$1$}\\ 
\hline
\end{tabular}
\caption{The charge assignment of the fields of the $\tilde{T}^{[3,2]}[SU(5)]$ theory, as seen from the quiver in figure \ref{(1)(3)[5]}.\label{table(1)(3)[5]}}
\end{table} 

Furthermore, from the quiver we can write down the superpotential, and by writing the F-term equations we see that there is one relation in the adjoint of $SU(3)$ and two relations which are singlets under the gauge group. In particular the prefactor in this example takes the following form
\begin{equation}
\mbox{Pfc}(w_3,w_4,\tilde{t})=\PE\left[[1,1]_{w_3,w_4}\tilde{t}^2+2\tilde{t}^2\right]^{-1}.
\end{equation}

With this information, we can write the Hilbert series, which in this case is
\begin{equation}
\begin{aligned}
HS\left(\tilde{t},x\right)&=\int d\mu_{U(1)\times U(2)}\ \ \mbox{Pfc}(w_3,\tilde{t})\ \cdot\\
&\cdot \PE\left[w_{1}w_2^{-1}[1,0]_{w_3,w_4}\tilde{t}+w_{1}^{-1}w_2[0,1]_{w_3,w_4}\tilde{t}+\right.\\
&\left.+w_2[0,1]_{w_3,w_4}[0,0,0,1]_x\tilde{t}+w_2^{-1}[1,0]_{w_3,w_4}[1,0,0,0]_x\tilde{t}\right],
\end{aligned}
\end{equation}
where we recall that the Haar measure for a $U(1)\times U(3)$ gauge group is given by 
\begin{equation}
\begin{aligned}
\int d\mu_{U(1)\times U(3)}=\dfrac{1}{(2\pi i)^4}\oint_{|w_1|=1}\oint_{|w_2|=1}\oint_{|w_3|=1}\oint_{|w_4|=1}&\dfrac{dw_1}{w_{1}}\dfrac{dw_2}{w_{2}}\dfrac{dw_3}{w_{3}}\dfrac{dw_4}{w_{4}}\\
&\left(1-w_3w_4\right)\left(1-\dfrac{w_3^2}{w_4}\right)\left(1-\dfrac{w_4^2}{w_3}\right).
\end{aligned}
\end{equation}
By performing this computation explicitly, and expanding to low order in $t=\tilde{t}^2$ we find
\begin{equation}
HS(t,x)=1+[1,0,0,1]_xt+\left([2,0,0,2]_x+[1,0,0,1]_x+[0,1,1,0]_x\right)t^2+\mathcal{O}(t^3). \label{HS32fromHiggs}
\end{equation}

\subsubsection*{Matching with the dual Coulomb branch part of $[2,2,1]$}

We then again compare the result \eqref{HS32fromHiggs} with the result by using the 3d mirror symmetry and the restriction rule. The dual to the partition $[3,2]$ is $[2,2,1]$. The Hilbert series of the Coulomb branch part $\mathcal{C}_{[2,2,1]}$ is given by 
\begin{equation}
HS(t,z)=1+[1,0,0,1]_zt+\left([2,0,0,2]_z+[1,0,0,1]_z+[0,1,1,0]_z\right)t^2+\mathcal{O}(t^3),
\end{equation}
which again yields the perfect agreement with \eqref{HS32fromHiggs}. This matching has been checked up to order $4$ in $t$. The relations between $z$ and $x$ are 
\begin{equation}
z_1 \mapsto x_1^{2}x_2^{-1}, \quad z_2\mapsto x_1^{-1}x_2^{2}x_3^{-1}, \quad z_3\mapsto x_2^{-1}x_3^2x_4^{-1}, \quad z_4\mapsto x_3^{-1}x_4^{2},
\end{equation}
since the Cartan matrix of the $su(5)$ Lie algebra is 
\begin{equation}
M_{A_4}=\left(\begin{array}{cccc}
2 & -1 & 0 & 0 \\ 
-1 & 2 & -1 & 0 \\ 
0 & -1 & 2 & -1 \\ 
0 & 0 & -1 & 2
\end{array} \right).
\end{equation}

\section{Conclusion and discussion}
\label{sec:conclusion}

In this paper we have determined the restriction rule for computing the Hilbert series for the Coulomb branch part of a mixed branch of the 3d $\mathcal{N}=4$ $T[SU(N)]$ theory from the Hilbert series of the full Coulomb branch. In particular, the brane realization of the mixed branch precisely gives an explicit way to truncate the magnetic charges as well as to reduce the classical dressing factor. We confirmed the method by comparing the result obtained by the restriction with the result obtained by the technique after going to the IR gauge theory.

We also computed the Hilbert series of the Higgs branch part of a mixed branch of the 3d $T[SU(N)]$ theory in two ways. One way is to use the technique by making use of the Molien-Weyl projection discussed in \ref{sec:HSHiggs}. In order to use the method, we consider an IR gauge theory by decoupling the Coulomb branch moduli. With the IR theory, we can apply the method and were able to compute the Hilbert series of the Higgs branch part. The other way is to utilize the 3d mirror symmetry and the restriction rule for computing the Coulomb branch part of a mixed branch. Intriguingly, the completely different computation exactly gives the same result including flavor fugacities. This provides a non-trivial check of the restriction rule as well as the mirror symmetry of the 3d $\mathcal{N}=4$ theories. 

By taking the product of the Hilbert series of the two branches, we are able to compute the Hilbert series of a mixed branch of the $T[SU(N)]$ theory. The restriction rule indeed gives a systematic way to obtain the series from the product of the Hilbert series of the two full branches.

Although our computation determines the Hilbert series of a mixed branch of the 3d $T[SU(N)]$ theory from the restriction rule, it is interesting to consider the Hilbert series of the full moduli space of the 3d $T[SU(N)]$ theory. In fact, the restriction procedure seems to suggest a natural way to obtain it. The basic structure of the Hilbert series of the full Coulomb branch moduli space of the $T[SU(N)]$ theory is that it is given by a sum of a set of magnetic charges $ \{\vec{m}\}$ as\footnote{We suppress the flavor fugacities for simplicity.}
\be
HS\left(t\right) = \sum_{\{\vec{m}\}}f\left(\{\vec{m}\}, t\right).  \label{HSfullCB}
\ee
The restriction rule says that among the possible summation of $\{\vec{m}\}$, there are special sub-summations where a Higgs branch opens up. For example, when $\vec{m}=\vec{0}$, which implies the origin of the Coulomb branch moduli space, we have the full Higgs branch which shares the origin. There is a natural guess to implement the intersection to the Hilbert series. Namely from the summation \eqref{HSfullCB}, we remove $\vec{m} = \vec{0}$, and add the Hilbert series of the full Higgs branch which we denote by $HS_{\mathcal{H}_{[N]}}(t)$,
\be
HS\left(t\right)  = \sum_{\{\vec{m}\}\setminus \{\vec{0}\}}f\left(\{\vec{m}\}, t\right) + HS_{\mathcal{H}_{[N]}}(t).
\ee
This guess will also lead to a way of incorporating another mixed branch further. For example, there is a mixed branch specified by a partition $[N-1, 1]$. The restriction rule says that for computing the Hilbert series of the Coulomb branch $\mathcal{C}_{[N-1, 1]}$, we sum over a subset of $\{\vec{m}\}$ and we denote the subset by $\{\vec{m}\}|_{R_{[N, 1]}}$, which also includes the origin. The $R_{[N-1,1]}$ is the restriction map introduced in section \ref{sec:restriction}. The Hilbert series of the Coulomb branch part can be written by $HS_{\mathcal{C}_{[N-1,1]}}(t) = \sum_{\{\vec{m}\}|_{R_{[N, 1]}}}f(\{\vec{m}\}, t)|_{R_{[N-1,1]}}$. Along the sublocus, a Higgs branch $\mathcal{H}_{[N, 1]}$ opens up and we denote the Hilbert series for $\mathcal{H}_{[N-1,1]}$ by $HS_{\mathcal{H}_{[N-1,1]}}(t)$. Then the Hilbert series with the mixed branch might be
\bea
HS\left(t\right) &=&  \sum_{\{\vec{m}\}\setminus \{\vec{m}\}|_{R_{[N-1,1]}}}f\left(\{\vec{m}\}, t\right)\nonumber\\
 &&+ \left(\sum_{\{\vec{m}\}|_{R_{[N-1,1]}}\setminus \{\vec{m}\}|_{R_{[N]}}}f\left(\{\vec{m}\}, t\right)|_{R_{[N-1,1]}}\right)\times HS_{\mathcal{H}_{[N-1,1]}}(t) \nonumber\\
&&+ HS_{\mathcal{H}_{[N]}}(t),
\eea
where $\{\vec{m}\}|_{R_{[N]}} = \vec{0}$. Therefore, the restriction rule yields a natural guess of computing the Hilbert series of the full moduli space by removing some magnetic charges corresponding to a sublocus and adding the Hilbert series of the mixed branch which stems from the sublocus. The repetition of the procedures would give a systematic way to compute the Hilbert series of the full moduli space of the 3d $\mathcal{N}=4$ $T[SU(N)]$ theory although the combinatorics of dividing the summation will be more complicated. At least, we checked the above procedure is consistent with the Hilbert series of a variety made from two $\mathbb{C}^n$-planes glued at a point. A similar gluing was first discussed in \cite{Hanany:2006uc}. It would be certainly interesting to prove this guess and we leave it for future work.

We hope the result obtained in this paper could be useful for future studies on the mixed branches of the moduli space of more general $3d$ $\mathcal{N}=4$ supersymmetric theories. One interesting direction of the generalization is to include $O3^{\pm}$-planes to the brane picture, and therefore to determine the restriction rule for computing the Hilbert series of mixed branches of the $T[SO(N)]$ and $T[Sp(N)]$ theories constructed in \cite{Gaiotto:2008ak}.

\bigskip

\bigskip

\centerline{\bf \large Acknowledgments}
\vspace{0.5cm}
F.C. and H.H. would like to thank Amihay Hanany and Stefano Cremonesi for comments on the manuscript and fuitful discussions.
F.C would like to thank the ICTP for kind hospitality during the workshop on ``Geometric correspondence of gauge theories 2016", when this paper was finalized. The work of F.C. is supported through a fellowship of the international programme ``La Caixa-Severo Ochoa". The work has been partially supported by the ERC Advanced Grant SPLE under contract ERC-2012-ADG-20120216-320421, by the grant FPA2012-32828 from the MINECO, and by the grant SEV-2012-0249 of the “Centro de Excelencia Severo Ochoa” Programme.

\bibliographystyle{JHEP}
\bibliography{refs}

\end{document}